\documentstyle[11pt,epsf]{article}

\newcommand{\be}{\begin{equation}}
\newcommand{\ee}{\end{equation}}
\newcommand{\nn}{\nonumber}
\newcommand{\beba}{\begin{equation}\begin{array}{lcl}}
\newcommand{\eaee}{\end{array}\end{equation}}
\newcommand{\bea}{\begin{eqnarray}}
\newcommand{\eea}{\end{eqnarray}}
\newcommand{\ba}{\begin{array}}
\newcommand{\ea}{\end{array}}

\newcommand{\ns}{\normalsize}
\newcommand{\refs}[1]{(\ref{#1})}

\def\abar{{\bar{a}}}
\def\bbar{{\bar{b}}}
\def\cbar{{\bar{c}}}

\def\Ib{{\bar{I}}}
\def\Jb{{\bar{J}}}
\def\Kb{{\bar{K}}}
\def\Lb{{\bar{L}}}

\def\tr{{\rm tr}}

\def\agut{\alpha_{\rm GUT}}

\def\Fd{\tilde{F}}
\def\Rd{\tilde{R}}
\def\RwR{{{\rm tr}R\wedge R}}
\def\FwFa{{{\rm tr}F^{(1)}\wedge F^{(1)}}}
\def\FwFb{{{\rm tr}F^{(2)}\wedge F^{(2)}}}

\newcommand{\cyav}[1]{\big<{#1}\big>_{\rm CY}}
\newcommand{\orbav}[1]{\big<{#1}\big>_{11}}
\newcommand{\av}[1]{\big<{#1}\big>_{\rm CY,11}}

\def\a{\alpha}
\def\b{\beta}
\def\g{\gamma}
\def\c{\chi}
\def\d{\delta}
\def\e{\epsilon}

\def\k{\kappa}
\def\l{\lambda}
\def\m{\mu}
\def\n{\nu}
\def\o{\omega}
\def\p{\pi}
\def\q{\theta}
\def\th{\theta}

\def\r{\rho}
\def\s{\sigma}

\def\x{\xi}

\def\D{\Delta}

\def\G{\Gamma}
\def\J{\Psi}

\def\O{\Omega}

\def\cB{{\cal B}}
\def\cH{{\cal H}}

\begin{document}

%%%%%%%%%%%%%%%%%%%%%%%%%%%%%%%%%%%%%%%%%%%%%%%%%%%%%%%%%%%%%%%%%%%%%%%%%%%

\begin{titlepage}
\vspace{-2cm}
\title{\hfill{\ns UPR-771T, PUPT-1723, HUB-EP-97/51\\}
       \hfill{\ns hep-th/9710208\\[.5cm]}
       {\Large\bf On the Four-Dimensional Effective Action of Strongly 
          Coupled Heterotic String Theory}}
\author{Andr\'e
        Lukas$^1$\setcounter{footnote}{0}\thanks{Supported by Deutsche
        Forschungsgemeinschaft (DFG) and
        Nato Collaborative Research Grant CRG.~940784.}~~,
        Burt A.~Ovrut$^1\, ^3\,$\setcounter{footnote}{3}\thanks{Supported
        in part by a Senior Alexander von Humboldt Award}~~and Daniel
         Waldram$^2$\\[0.5cm]
        {\ns $^1$Department of Physics, University of Pennsylvania} \\
        {\ns Philadelphia, PA 19104--6396, USA}\\[0.3cm]
        {\ns $^2$Department of Physics}\\
        {\ns Joseph Henry Laboratories, Princeton University}\\
        {\ns Princeton, NJ 08544, USA}\\[0.3cm]
        {\ns $^3$ Institut f\"ur Physik, Humboldt Universit\"at}\\
        {\ns Invalidenstra\ss{}e 110, 10115 Berlin, Germany}}
\date{}
\maketitle

\begin{abstract}
The low-energy $D=4$, $N=1$ effective action of the strongly coupled
heterotic string is explicitly computed by compactifying Ho\v rava-Witten
theory on the deformed Calabi-Yau three-fold solution due to Witten.
It is shown that, to order $\k^{2/3}$, the K\"ahler potential is identical
to that of the weakly coupled theory. Furthermore, the gauge kinetic
functions are directly computed to order $\k^{4/3}$ and shown to receive
a non-vanishing correction. Also, we compute gauge matter terms in the
K\"ahler potential to the order $\k^{4/3}$ and find a nontrivial correction
to the dilaton term. Part of those corrections arise from background fields
that depend on the orbifold coordinate and are excited by four-dimensional
gauge field source terms.

\end{abstract}

\thispagestyle{empty}
\end{titlepage}

%%%%%%%%%%%%%%%%%%

\section{Introduction}

The simplest evidence that models of particle physics might best be
described by the strong-coupling limit of heterotic string theory
comes from the string tree-level predictions of the four-dimensional
Planck scale and the grand unification scale and
coupling. Generically, it is
difficult to match the experimentally inferred values
without a string coupling many orders of magnitude larger than
unity. Witten~\cite{w} has investigated this possibility in the case of the
$E_8\times E_8$ heterotic string, using the identification of the
strongly coupled limit with M theory compactified on an $S^1/Z_2$
orbifold~\cite{hw1,hw2}. The effective action of M theory can be
formulated as a momentum expansion in terms of powers of the
eleven-dimensional gravitational coupling $\k^2$. Remarkably, Witten
finds a model where the four-dimensional Planck scale and the inferred
grand unification scale and coupling can just be matched, without
corrections in the expansion becoming too large~\cite{w,bd}. In addition,
Ho\v rava has shown that in this model supersymmetry can be broken via 
an interesting topological version of gluino condensation~\cite{hor}. 

The purpose of the present paper is to derive, directly from M theory,
the $N=1$ four-dimensional effective action which arises from Witten's strongly
coupled model. As a preparation, we prove several useful properties of
Witten's solution. In particular, we introduce a simple reparameterization
of this solution and find its explicit form in terms of a harmonic expansion.
Armed with these results, we calculate the corrections due to
terms of higher order in $\k^2$, which are the analog of loop-corrections
in the weakly coupled string. We present the K\"ahler potential to order
$\k^{2/3}$ for the moduli fields and to order $\k^{4/3}$ in the
gauge matter fields. For the gauge fields we calculate the gauge kinetic
functions to order $\k^{4/3}$. 
%Apparently, a related calculation for the
%real part of the gauge threshold correction was carried out in ref.~\cite{hp},
%although details and the exact coefficient were not presented.
In general, we take care to identify contributions from background fields
excited by four--dimensional gauge and gauge matter field sources.
We find these background fields contribute to various corrections
at order $\k^{4/3}$. In particular, they generate the imaginary part of the
threshold, with the real part arising from the metric distortion due
to the internal gauge fields. For the K\"ahler potential, we show
that it is possible to choose a parameterization of the universal
moduli, such that their K\"ahler potential receives no
corrections to the order $\k^{2/3}$. To the order $\k^{4/3}$,
we find a gauge matter field term correcting the dilaton part of the
K\"ahler potential, which is the direct analog of the order $\k^{4/3}$
threshold corrections to the gauge kinetic functions.

Some of these results have been anticipated
in the literature. The K\"ahler potential, to zeroth order in $\k$,
was computed in ref.~\cite{du1,aq,lt,du2} but these authors did not discuss
possible order $\k^{2/3}$ modifications. Such modifications, and their
absence to this order, were argued for indirectly in ref.~\cite{bd}. 
The modification of the gauge coupling was first discussed by Witten
in his presentation of the strongly coupled model. Arguments for the form
of the gauge kinetic functions, based on results in the weakly
coupled limit, symmetry and holomorphicity, were presented in
ref.~\cite{bd,ns,choi}. 
It appears that a related calculation to the
one presented here, for the real part of the gauge threshold
correction, was carried out in ref.~\cite{hp}, although details and
the exact coefficient were not presented. 
Phenomenological aspects of the zeroth order effective
action of the strongly coupled heterotic string have been discussed in
ref.~\cite{lln,choi,efn}. Gaugino condensation in this model has been
considered in ref.~\cite{hor,hp,lt}

\vspace{0.4cm}

It is useful to understand the scales in Witten's model and also
the exact nature of the expansion in $\k^2$. To do so, let us be a
little more explicit about the Ho\v rava-Witten description of the
strongly coupled heterotic string, and also about Witten's background
solution.  To zeroth order in a momentum expansion in $\k^2$, M theory is
simply eleven-dimensional supergravity~\cite{D11}. When compactified on
$S^1/Z_2$,  the first correction appears at order $\k^{2/3}$. One finds that
a set of $E_8$ gauge fields must be included on each of the two fixed
hyperplanes of the orbifold and the Bianchi identity for the
supergravity four-form field $G$ gets a correction. These terms are
required if the theory is to be free of anomalies. 

To make contact with low-energy physics, Witten considered a
compactification which, to zeroth order in $\k^2$, was simply the
$S^1/Z_2$ orbifold times a Calabi-Yau three-fold. Two
universal moduli appear, which are the analogs of the $S$ and $T$
moduli of the weakly coupled limit, namely, the volume of the
Calabi-Yau three-fold $V$ and the radius of the orbifold $\rho$. 
One finds that the four-dimensional Newton constant and the
grand-unified (GUT) coupling constant are given by 
\be 
  G_N = \frac{\k^2}{16\pi^2V\rho} \qquad
  \agut = \frac{\left(4\pi\k^2\right)^{2/3}}{2V} \; , \label{couplings}
\ee
while the GUT scale (about $10^{16}$ Gev) is set by the size of the
Calabi-Yau manifold $V^{1/6}$. To match the known values of $G_N$ and
$\agut$, Banks and Dine~\cite{bd} find that $V^{1/6}/\k^{2/9}\sim 2$ while
$\pi\rho/\k^{2/9}\sim 8$. That is to say, the Calabi-Yau radius is
about twice as large as the eleven-dimensional Planck scale
$\k^{2/9}$, while the orbifold radius is about an order of magnitude
larger. (These numbers can vary somewhat depending on how exactly the
scales are defined.) Thus in such a scenario, with increasing energy, the
universe appears first five- and then eleven-dimensional, and the
four-dimensional Planck scale is relegated to a low-energy parameter,
with no direct relevance to the scale of quantum gravitational effects. 

To this order, other than the difference in scales, the low-energy
theory is indistinguishable from that for the tree-level weakly coupled
string. However, Witten calculates the modification in the M theory
background due to the next order corrections in $\k^2$, which appear
at $\k^{2/3}$, and which still preserves $N=1$ supersymmetry in four
dimensions. In analogy to the weakly coupled case, the spin connection
is embedded in one of the $E_8$ groups. However, the supergravity
four-form can no longer be taken to be zero because of the correction
to its Bianchi identity. This in turn means that the internal space
becomes deformed and is no longer a simple product of the $S^1/Z_2$
orbifold with a Calabi-Yau three-fold. It is the modification to the
low-energy action due to this deformation in which we are
interested.  

To understand the form of potential corrections to the low-energy
action, one needs to identify what exactly is the expansion
parameter in Witten's solution. Since 
there are two dimensionless scales, related to the size of the
Calabi-Yau space and the orbifold in eleven-dimensional Planck units, we do
not {\em a priori} know if the expansion is in $\k^{2/3}/V^{1/2}$ or
$\k^{2/3}/\rho^3$ or some combination thereof. In fact, one finds that
the metric deformation in Witten's solution is of order
$\e=\k^{2/3}\rho/V^{2/3}$. Using the Banks and Dine values for $V$ and
$\rho$ given above, one finds $\e\sim\frac{1}{2}$, and we are right at the
limit of the validity of the expansion. However, partly since $\e$
depends somewhat on exactly how the scales are defined, we shall
assume in this paper that the expansion is reasonable. From the form
of $\e$, we find that the first-order corrections to the low-energy
action should depend on the universal moduli in the combination
$\rho/V^{2/3}$. 
%In terms of the $S$ and $T$ moduli of the weakly
%coupled limit our expansion can be summerized as follows. Both the
%Calabi-Yau and orbifold sizes are taken to be large compared to the
%eleven-dimensional Planck scale. Thus we require $S\gg 1$ and
%$T\gg 1$. Meanwhile, the solution is an expansion in $\e\ll 1$, which is
%equivalent to the requirement $T/S\ll 1$. As we have already
%mentioned, matching the gravitational and grand-unified couplings
%gives $S\sim 60$ and $T \sim 30$ so we are right on the edge for a good
%expansion in $\e$. 

Let us conclude this section by summarizing our conventions. 
We denote the eleven-dimensional coordinates
by $x^0,\ldots,x^9,x^{11}$ and the corresponding indices by
$I,J,K,\ldots=0,\ldots,9,11$. The orbifold  $S^1/Z_2$ is chosen in the
$x^{11}$--direction, so we assume that  $x^{11}\in [-\pi\r ,\pi\r ]$ with
the endpoints identified as $x^{11}\sim x^{11}+2\pi\r$. The $Z_2$ symmetry
acts as $x^{11}\rightarrow -x^{11}$. Then there exist two ten-dimensional
hyperplanes, $M^{10}_i$ with $i=1,2$, locally specified by the
conditions $x^{11}=0$ and $x^{11}=\pi\r$, which are fixed under the
action of the $Z_2$ symmetry. We will sometimes use the ``downstairs''
picture where the orbifold is considered as an interval 
$x^{11}\in [0,\pi\r ]$ with the fixed hyperplanes forming boundaries to the
eleven-dimensional space. In the ``upstairs'' picture the eleventh
coordinate is considered as the full circle with singular points at the
fixed hyperplanes. We will use barred indices 
$\bar{I},\bar{J},\bar{K},\ldots=0,\ldots,9$ to label the ten-dimensional 
coordinates. When we later further compactify the theory on a
Calabi-Yau three-fold, we will use indices $A,B,C,\ldots=4,\ldots,9$
for the Calabi-Yau coordinates, and indices $\mu,\nu\ldots=0,\ldots,3$
for the coordinates of the remaining, uncompactified, four-dimensional
space. Holomorphic and antiholomorphic coordinates on the Calabi-Yau space 
will be labeled by $a,b,c,\ldots$ and $\abar,\bbar,\cbar,\ldots$.

\section{The strongly coupled heterotic string and compactification on 
Calabi--Yau spaces}

In the first part of this section, we review the formulation of the
low-energy limit of strongly coupled heterotic string theory as the
effective action for M theory compactified on the orbifold $S^1/Z_2$.
The effective theory is given as a momentum expansion in the 
eleven-dimensional gravitational coupling $\k^{2/3}$. As an example of
techniques we will use in deriving effective actions in
four-dimensions, we then consider 
the limit where the fields are assumed to vary slowly over the scale
of the orbifold interval. The resulting ten-dimensional theory, though
calculated in the strongly coupled limit, is identical to the
low-energy effective action for the weakly coupled heterotic
string. In particular, we demonstrate how the ten-dimensional
Chern-Simons and Green-Schwarz terms appear through a somewhat novel
mechanism, arising from components of the supergravity four-form which
depend non-trivially on the orbifold dimension. This mechanism will
reappear when we come to derive the effective action in four
dimensions. Finally, we review the form of a solution of the
eleven-dimensional theory, due to Witten, which has a spacetime
geometry $M^{11}=S^1/Z_2\times\ X\times M^4$, where $X$ is a deformed
Calabi--Yau three-fold and $M^4$ is the Minkowski four-space. This
background is the starting point for the construction of a $D=4$
effective action with $N=1$ supersymmetry. 

\subsection{M theory on $S^1/Z_2$}

To zeroth order in $\k^2$, the effective action for M theory on $S^1/Z_2$ is 
simply that of eleven-dimensional supergravity. In the upstairs
picture, the bosonic parts are given by 
\be
 S_{\rm SG} = \frac{1}{2\k^2}\int_{M^{11}}\sqrt{-g}\left[ 
                    -R-\frac{1}{24}G_{IJKL}G^{IJKL}
           -\frac{\sqrt{2}}{1728}\e^{I_1...I_{11}}
               C_{I_1I_2I_3}G_{I_4...I_7}G_{I_8...I_{11}} \right] \; .
 \label{LSG}
\ee
The field $C_{IJK}$ is the three-form potential for the field strength
$G_{IJKL}=24\partial_{[I}C_{JKL]}$. In addition, the fields are 
restricted to respect the $Z_2$ orbifold
symmetry. For the bosonic fields, $g_{\bar{I}\bar{J}}$, $g_{11\, 11}$
and $C_{11\bar{I}\bar{J}}$ must be even under $Z_2$, while
$g_{11\bar{I}}$ and $C_{\bar{I}\bar{J}\bar{K}}$ must be odd. For
the eleven dimensional gravitino the condition is
\be
  \J_{\bar{I}}(x^{11}) = \G_{11}\J_{\bar{I}}(-x^{11}) \qquad 
  \J_{11}(x^{11}) = -\G_{11}\J_{11}(-x^{11}) \; .
\ee
This constraint means that the gravitino is chiral from a
ten-dimensional perspective and so, unlike the usual case in
eleven-dimensions, the theory has a gravitational anomaly, localized
on the fixed hyperplanes. 

The way to cancel this anomaly is to introduce two
ten-dimensional $E_8$ gauge supermultiplets, one on each of the two
fixed hyperplanes of the orbifold. The leading correction to the
supergravity action~\refs{LSG} enters at order $\k^{2/3}$. If we restrict 
ourselves for the moment to terms at most quadratic in derivatives, one 
finds the action is given by
\bea
  S &=& S_{\rm SG}+S_{\rm YM} \nn \\
    &=& S_{\rm SG} 
            - \frac{1}{8\pi\k^2}\left(\frac{\k}{4\pi}\right)^{2/3}
               \int_{M^{10}_1}\sqrt{-g}\;{\rm tr}(F^{(1)})^2
            -  \frac{1}{8\pi\k^2}\left(\frac{\k}{4\pi}\right)^{2/3}
               \int_{M^{10}_2}\sqrt{-g}\;{\rm tr}(F^{(2)})^2
  \label{LYM}
\eea
where $F_{\bar{I}\bar{J}}^{(1,2)}$ are the field strengths of the two
$E_8$ gauge fields $A_{\bar{I}}^{(1,2)}$. Hidden in the above expression is
a further order $\k^{2/3}$ correction. Namely, to preserve supersymmetry,
it is necessary to correct the Bianchi identity for $G_{IJKL}$, introducing
source terms localized on the hyperplanes so that
\be
 (dG)_{11\bar{I}\bar{J}\bar{K}\bar{L}} = -\frac{1}{2\sqrt{2}\pi}
    \left(\frac{\k}{4\pi}\right)^{2/3} \left\{ 
       J^{(1)}_{\bar{I}\bar{J}\bar{K}\bar{L}} \d (x^{11})
       + J^{(2)}_{\bar{I}\bar{J}\bar{K}\bar{L}} \d (x^{11}-\pi\r )
       \right\} \; .\label{Bianchi}
\ee
where the sources are given by 
\bea
 J^{(i)}_{\bar{I}\bar{J}\bar{K}\bar{L}} 
    &=& \left( {\rm tr}F^{(i)}\wedge F^{(i)} 
          - \frac{1}{2}{\rm tr}R\wedge R \right)_{\Ib\Jb\Kb\Lb} \nn \\
    &=& 6\left( {\rm tr}F^{(i)}_{[\bar{I}\bar{J}}F^{(i)}_{\bar{K}\bar{L}]}
         - \frac{1}{2}{\rm tr}R_{[\bar{I}\bar{J}}R_{\bar{K}\bar{L}]}
      \right) \; . \label{Jdef}
\eea
Note, that the gravitational contribution to this Bianchi
identity is distributed equally between the two
hyperplanes, giving rise to the factor $1/2$ in front of tr$R^2$. 
As Ho\v rava and Witten point out~\cite{hw2}, if $G$ is to be free
from delta functions, this Bianchi identity is equivalent to a
boundary condition on the value of $G$ at the two fixed
hyperplanes. One finds 
\bea
 \left.G_{\bar{I}\bar{J}\bar{K}\bar{L}}\right|_{x^{11}=0} 
   &=& -\frac{1}{4\sqrt{2}\pi}\left(\k/4\pi\right)^{2/3} 
        J^{(1)}_{\bar{I}\bar{J}\bar{K}\bar{L}} \nn \\
 \left.G_{\bar{I}\bar{J}\bar{K}\bar{L}}\right|_{x^{11}=\pi\r} 
   &=& \frac{1}{4\sqrt{2}\pi}\left(\k/4\pi\right)^{2/3} 
        J^{(2)}_{\bar{I}\bar{J}\bar{K}\bar{L}} \label{bdryG}
\eea
at the two boundaries.

In what follows, we will see that the action must also have additional terms 
at order $\k^{2/3}$ which are higher-order in derivatives. In particular, we 
will find evidence for terms quadratic in the curvature localized on the two 
hyperplanes. These enter the action as the Gauss-Bonnet combination, 
\be
 S_{R^2} = \frac{1}{16\pi\k^2}\left(\frac{\k}{4\pi}\right)^{2/3}
      \int_{M^{10}_1,M^{10}_2}\sqrt{-g}\;\left(
           R_{\bar{I}\bar{J}\bar{K}\bar{L}}R^{\bar{I}\bar{J}\bar{K}\bar{L}}
           - 4R_{\bar{I}\bar{J}}R^{\bar{I}\bar{J}} + R^2 \right) \; ,
      \label{SR2}
\ee
where the integral is over both hyperplanes. (In fact, we will not be
able to fix the coefficients of the Ricci tensor and scalar terms
in~\refs{SR2}. This is typically also the case in the calculating
perturbative string effective actions. As is often done, we will
assume it is the Gauss-Bonnet combination which appears since this is
free of ghosts.)

Away from the hyperplanes, the theory at order $\k^{2/3}$
is still simple eleven-dimensional supergravity. Including the contributions
from the hyperplanes, the Einstein equation gets additional source terms
localized on the boundaries. The corresponding equations of motion are
then 
\bea
 D_I G^{IJKL} &=& \frac{\sqrt{2}}{1152}\e^{JKLI_1 \ldots I_8} 
    G_{I_1 \ldots I_4}G_{I_5 \ldots I_8} \nn \\ \label{Geom}
 R_{IJ} - \frac{1}{2}g_{IJ}R &=& -\frac{1}{24} \left(4G_{IKLM}{G_J}^{KLM} - 
    \frac{1}{2}g_{IJ}G_{KLMN}G^{KLMN}\right)\nn \\
 &&-\frac{1}{2\pi}(\k /4\pi )^{2/3}\left(\d (x^{11})T_{IJ}^{(1)}
   +\d (x^{11}-\pi\r )T_{IJ}^{(2)}\right) \label{geom} \; ,
\eea
where $T^{(i)}_{IJ}$ is the energy momentum tensor of the gauge field
$A^{(i)}$ plus the contribution from the $R^2$ terms~\refs{SR2}.
Concentrating on the gauge fields, its only nonvanishing components
are given by
\be
 T^{(i)}_{\bar{I}\bar{J}} =(g_{11,11})^{-1/2}\left(
       \mbox{tr}F_{\bar{I}\bar{K}}^{(i)}F_{\bar{J}}^{(i)\bar{K}}
       - \frac{1}{4}g_{\bar{I}\bar{J}}\mbox{tr}(F^{(i)})^2\right)\; .
 \label{emt}
\ee
The supersymmetry transformation of the gravitino $\J_I$ to this
order in $\k$ is unchanged from the supergravity expression, 
\be 
 \d\J_I = D_I\eta +\frac{\sqrt{2}}{288}\left(\G_{IJKLM}-8g_{IJ}\G_{KLM}
          \right)G^{JKLM}\eta +\; \cdots \label{susy}\; ,
\ee
where the dots indicate the omitted fermionic terms and $\eta$ is an
eleven-dimensional Majorana spinor. This spinor should be restricted by the
condition
\be
 \eta (x^{11}) = \G_{11}\eta (-x^{11}) \label{Z2eta}
\ee
for the supersymmetry variation~\refs{susy} to be compatible with the
$Z_2$ symmetry. This constraint means that the theory has the usual 32
supersymmetries in the bulk but only 16 (chiral) supersymmetries on the
10--dimensional orbifold hyperplanes.

While some terms were calculated by Ho\v rava and Witten~\cite{hw2}, the
full M theory action to the next order in expansion, namely $\k^{4/3}$,
is not known. In general, this will mean that we will not trust 
background solutions or correction terms in a four-dimensional effective
action beyond $\k^{2/3}$. However, we will show that certain terms of
order $\k^{4/3}$, namely those involving gauge or gauge matter fields,
can be reliably computed.

\subsection{The ten-dimensional limit}

One can ask what form the low-energy effective action of the previous section 
takes when reduced to an effective ten-dimensional theory. That is, when one 
considers fields which vary slowly on the scale of the size of the orbifold 
interval. This is a different limit from the one we will consider 
in deriving the effective four-dimensional action. As we
discussed  in the introduction, in the latter case the theory is further
compactified on a 
six-dimensional Calabi-Yau manifold, where the size of the Calabi-Yau space is 
smaller than the size of the orbifold. Thus there is no scale at which 
the universe appears ten-dimensional. However, going to the ten-dimensional 
limit does highlight many of the particular techniques we will use in deriving 
an effective action in four-dimensions. A more detailed discussion of
this procedure will appear in \cite{low}.

It is also interesting to compare the ten-dimensional limit to the
weakly coupled theory. The ten-dimensional theory derived in 
this subsection describes the effective M-theory action for fields with 
momenta below the scale set by the orbifold interval. The interval is still 
assumed to be large compared with the eleven-dimensional Planck length and 
so the theory still describes the strongly coupled string. Nonetheless, we 
will find that the low-energy theory exactly reproduces the one-loop effective 
action of the weakly coupled heterotic string. This is, in fact, not surprising 
since this form is completely fixed by anomaly cancellation and
supersymmetry. It is essentially the statement that the
form of the lowest-dimension terms of the low-energy weakly coupled
action cannot get corrections. 
This serves as simple evidence that the M-theory action describes
the low-energy strongly coupled limit of the heterotic string. 
A related calculation was performed by Dudas and
Mourad~\cite{DM}. However, these authors made a projection onto one
hyperplane of the orbifold rather than a Kulaza-Klein
reduction, where one averages over the orbifold interval. As such,
they were unable to exactly identify the coefficients of the
Chern-Simons and Green-Schwarz terms. Here we find that the exact
ten-dimensional coefficients appear. 

The ten-dimensional theory arises as the familiar Kaluza-Klein
truncation of the eleven-dimen\-sional effective action. We have already
noted that only the $g_{{\bar I}{\bar J}}$ and $g_{11\,11}$ components
of the metric, together with the $C_{{\bar I}{\bar J}11}$ components of the
four-form field, are even under the orbifold $Z_2$ symmetry. Since the
massless Kaluza-Klein modes are independent of $x^{11}$, only these
components can survive in ten dimensions. The bosonic fields are thus
given by 
\be
 ds^2 = e^{-c/4}g_{\Ib\Jb}dx^\Ib dx^\Jb + e^{2c}(dx^{11})^2
 \qquad
 C_{\Ib\Jb 11} = \frac{1}{6}B_{\Ib\Jb} \; , \label{gBdef}
\ee
where the conformal factor in the ten-dimensional metric is chosen to
put the truncated action in canonical form. Reducing the zeroth-order
action~\refs{LSG}, we find the bosonic part of the ten-dimensional
theory is given by 
\be
 S_{10} = \frac{\pi\r}{\k^2}\int_{M^{10}}\sqrt{-g}\left\{- R 
           - \frac{9}{8}\left(\partial c\right)^2 
           - \frac{1}{6}e^{-3c/2}H^2 \right\} \; , \label{S10-0}
\ee
where $H_{\Ib\Jb\Kb}=3\partial_{[\Ib}B_{\Jb\Kb]}$. We immediately recognize 
equation~\refs{S10-0} as the action for ten-dimensional $N=1$ supergravity. 

Turning to the order $\k^{2/3}$ corrections one must now consider
gauge fields. Furthermore the source terms in the Bianchi
identity~\refs{Bianchi}  no longer vanish or, equivalently,
$G_{\Ib\Jb\Kb\Lb}$ is, in general, non-zero 
at the boundary. This leads to an interesting effect. To zeroth order,
since the field $G_{\Ib\Jb\Kb\Lb}$ is not even under the $Z_2$
symmetry, we expect it to be set to zero and play no role in the
reduced ten-dimensional theory. Now, however, in order to match the two
boundary conditions, $G_{\Ib\Jb\Kb\Lb}$ is not zero. In fact,
it must vary across the orbifold interval since, in general, the sources are
not equal. Thus, we find that, although we try to consider only fields
which depend on ten dimensions, exciting gauge fields or curvature
necessary leads to exciting a component of $G$ which depends on the
internal $x^{11}$ coordinate. 

We can try and calculate the effect of these sources by finding an
approximate solution for $G$. Consider the theory in the downstairs
picture. We can always write $G$ in terms of a potential $C$. However,
$G$ must also satisfy the boundary conditions~\refs{bdryG}, together
with the equation of motion~\refs{Geom}. Since the boundary conditions
imply $G$ is proportional to $\k^{2/3}$, working to first order in
$\k^2$, the equation of motion reduces to 
\be
  D^I G_{IJKL} = 0
\ee
In the limit we are considering, the orbifold
interval is much smaller than the scale on which the ten-dimensional
fields vary. Thus we can look for a solution as a momentum expansion
in derivatives of the boundary sources $J^{(i)}$. To do this, we first
note that the sources can be integrated as
\be
  J^{(i)} = \tr F^{(i)}\wedge F^{(i)} - \frac{1}{2}\tr R\wedge R
          = d \o^{(i)}_3 \; ,
\ee
where we define
\be
  \o^{(i)}_3 = \o^{{\rm YM},(i)}_3 - \frac{1}{2}\o^{\rm L}_3 \; ,
  \label{CS_def}
\ee
as the sum of the Yang-Mills and Lorentz Chern-Simons forms 
$\o^{{\rm YM},(i)}_3$ and $\o^{\rm L}_3$. The first-order solution is then
given by 
\be
  C_{\Ib\Jb\Kb} = -\frac{1}{24\sqrt{2}\pi}\left(\k/4\pi\right)^{2/3}
       \left\{ \o^{(1)}_3 
          - (x^{11}/\pi\r)(\o^{(1)}_3+\o^{(2)}_3)
          \right\}_{\Ib\Jb\Kb} \; , \label{Cbkgd}
\ee
where we are dropping correction terms of the form 
$\rho\partial J^{(i)}$ and higher as an expansion in derivatives. The
corresponding field strengths $G=6dC$ are then
\bea
  G_{\Ib\Jb\Kb\Lb} &=& -\frac{1}{4\sqrt{2}\pi}\left(\k/4\pi\right)^{2/3}
      \left\{ J^{(1)} - (x^{11}/\pi\r)(J^{(1)}+J^{(2)})
      \right\}_{\Ib\Jb\Kb\Lb} \nn \\
  G_{\Ib\Jb\Kb 11} &=& -\frac{1}{4\sqrt{2}\pi^2\r}\left(\k/4\pi\right)^{2/3}
      \left(\o^{(1)}_3+\o^{(2)}_3\right)_{\Ib\Jb\Kb} \; .
\eea
We see that the solution is a simple linear extrapolation between the
two values of $G_{\Ib\Jb\Kb\Lb}$ at the two boundaries. However, the
Bianchi identity links $G_{\Ib\Jb\Kb\Lb}$ to other components of $G$
and, in particular, we find $G_{\Ib\Jb\Kb 11}$ is non-zero. A similar
phenomenon has been discussed in the context of soliton solutions
in M--theory on $S^1/Z_2$~\cite{llo}. There, the full momentum expansion
including all higher derivative terms has been worked out for the
gauge five--brane solution.

In summary, at this order in $\k^2$, the boundary conditions on $G$
have forced us to include a non-zero $C_{\Ib\Jb\Kb}$, depending
linearly on $x^{11}$, which would have been set to zero in a typical
dimensional reduction. Similar $x^{11}$--dependent background fields
arise for the metric due to the gauge field stress energy on the
boundary. For a complete dimensional reduction, these metric backgrounds
have to be taken into account as well, and we will do so in the
reduction to four dimensions later on. In this subsection, however,
we will concentrate on the effect of the three--form background, which
already has a number of consequences. A complete reduction to
ten dimensions will be presented elsewhere~\cite{low}.

As before, the ten-dimensional fields are given by equation~\refs{gBdef}
together with the $E_8\times E_8$ gauge fields. However, the field
strength $G_{\Ib\Jb\Kb 11}$ now has two contributions
\bea
 G_{\Ib\Jb\Kb 11} \equiv H_{\Ib\Jb\Kb} &=& 3\partial_{[\Ib}B_{\Jb\Kb ]}
     - \frac{1}{4\sqrt{2}\pi^2\r}\left(\frac{\k}{4\pi}\right)^{2/3}
       \left\{\o^{(1)}_3+\o^{(2)}_3\right\}_{\Ib\Jb\Kb} \nn \\
    &=& 3\partial_{[\Ib}B_{\Jb\Kb ]}
     - \frac{1}{4\sqrt{2}\pi^2\r}\left(\frac{\k}{4\pi}\right)^{2/3}
       \left\{\o^{{\rm YM}(1)}_3+\o^{{\rm YM}(2)}_3-\o^{\rm L}_3
            \right\}_{\Ib\Jb\Kb} \; , \label{CSH}
\eea
where the new second term comes from the non-zero $C_{\Ib\Jb\Kb}$. We
see that the effect of this term is to generate exactly the same gauge
and Lorentz Chern-Simons terms that appear in the effective action for
the weakly coupled heterotic string. 

If we continue the dimensional reduction, including the gauge and
$R^2$ boundary terms, we find the ten-dimensional action to order
$\k^{2/3}$
\bea
 S_{10} &=& \frac{\pi\r}{\k^2}\int_{M^{10}}\sqrt{-g}\left\{ - R 
           - \frac{9}{8}\left(\partial c\right)^2 
           - \frac{1}{6}e^{-3c/2}H^2 
           - \frac{1}{8\pi^2\r}\left(\frac{\k}{4\pi}\right)^{2/3}e^{-3c/4}
                \left( \tr(F^{(1)})^2 + \tr(F^{(1)})^2 \right)
           \right. \nn \\ && \qquad\qquad\qquad \left.
           + \frac{1}{8\pi^2\r}\left(\frac{\k}{4\pi}\right)^{2/3}e^{-3c/4}
                \left( R_{\Ib\Jb\Kb\Lb}R^{\Ib\Jb\Kb\Lb}
                  - 4R_{\Ib\Jb}R^{\Ib\Jb} + R^2 \right) 
           \right\} \; , \label{S10-1}
\eea
where $H$ is the field strength with Chern-Simons terms as given in
equation~\refs{CSH}, and we have dropped higher-order terms with 
involving the scalar $c$. We see that we have exactly reproduced
the terms of the one-loop effective action for the weakly coupled
heterotic string. We note that the Gauss-Bonnet $R^2$ terms are
required by supersymmetry~\cite{susyR2}, pairing with the Lorentz
Chern-Simons terms in $H$. Unless such terms are included explicitly 
as boundary terms in the eleven-dimensional theory, they will not
appear upon dimensional reduction. (No bulk $R^2$ terms are allowed in
eleven-dimensional supergravity.) This provides our first evidence that
such terms are necessary. We shall see more evidence in the following
section. We expect that the presence of Gauss-Bonnet boundary terms
can also be demonstrated explicitly in eleven-dimensions by the
requirement of supersymmetry together with the boundary condition on
$G$, in a calculation similar to the calculation demonstrating their
presence in ten-dimensions~\cite{susyR2}. 

Finally, we come to the question of the Green-Schwarz terms. We know that
the ten-dimensional theory, as written, is anomalous without
Green-Schwarz terms. Ho\v rava and Witten have argued how the anomaly
is cancelled in eleven-dimensions, but how do the required terms appear
here? Concentrating on the gauge terms, as expected from the
eleven-dimensional argument, we find that they appear from the
reduction of the 
bulk $C\wedge G\wedge G$ in eleven-dimensional supergravity, entering
at order $\k^{4/3}$. Using the form of $C_{\Ib\Jb\Kb}$ given
in~\refs{Cbkgd} and the ten-dimensional fields~\refs{gBdef}, after
integrating over $x^{11}$, these terms contribute 
\be
  S_{\rm GS} = \frac{1}{12\sqrt{2}\pi^2}\frac{\pi\r}{\k^2}
     \left(\frac{\k}{4\pi}\right)^{4/3} \int_{M^{10}} B\wedge X^{\rm YM}_8
\ee
to the effective action, where 
\be
  X^{\rm YM}_8 = -\frac{1}{4}\left\{ \left(\tr{F^{(1)}}^2\right)^2 
         - \tr{F^{(1)}}^2\tr{F^{(2)}}^2
         + \left(\tr{F^{(2)}}^2\right)^2 \right\}
\ee
and wedge products are understood in this expression. We see that
$X_8$ is precisely the Yang-Mills contribution to the Green-Schwarz
term. This calculation can be extended to include the gravitational
contributions. These come partly from the dimensional reduction
of the $C\wedge G\wedge G$ term and partly from an explicit bulk M
theory term which couples $C$ to four curvature tensors. (Such a
term has been discussed in various places~\cite{CR4} and is necessary
for anomaly cancellation in the Ho\v rava-Witten formulation of the
strongly coupled heterotic string.) While we do not give the
details of either of the calculation here, it is possible to show
that through this mechanism, one reproduces exactly the full
Green-Schwarz term as appears in the effective action of the weakly
coupled heterotic string~\cite{low}.

We have seen that reducing the strongly coupled theory to ten
dimensions gives the same effective action as the weakly coupled
string to one-loop, including Chern-Simons and Green-Schwarz
terms. The appearance of these latter terms depended on the boundary
conditions on $G$ which led to a non-zero $C_{\Ib\Jb\Kb}$ depending on
$x^{11}$. In general, boundary sources for the Einstein equation means
that the metric also has components which depend non-trivially on
$x^{11}$, although we have not discussed them in detail here. In deriving
an effective action in four-dimensions, we will find an identical
mechanism leading to gauge field couplings at order $\k^{4/3}$. 

\subsection{Review of Witten's solution}

In order to make contact with low-energy physics, one would like to
consider compactifications of the strong-coupled theory which have $N=1$
supersymmetry in four dimensions. Such backgrounds have been constructed by
Witten~\cite{w}. To zeroth order the compactification is simple. The
$G$ equation of motion and zeroth-order Bianchi identity can be
satisfied by setting $G=0$. Supersymmetry is then preserved by
compactifying on a Calabi-Yau three-fold. The background metric then
has the form 
\be
 ds^2 = g^{(0)}_{IJ}dx^Idx^J
      = \eta_{\m\n} dx^\m dx^\n + \O_{AB}dx^Adx^B + (dx^{11})^2\; .
 \label{metric0}
\ee
$\O_{AB}$ is the metric of the Calabi-Yau space, which, in holomorphic
coordinates, is related to the K\"ahler form $\o_{AB}$ by 
$\o_{a\bar{b}}=-i\O_{a\bar{b}}$. As we have already mentioned, to
match to physical values of the grand-unified and gravitational
couplings and grand-unified scale, it is necessary to take the size of the
Calabi-Yau space smaller than the size of the orbifold
interval. Thus we are in a very different limit from the previous
subsection, where the orbifold was taken smaller than the scale of the
remaining ten dimensions.

What if we try to find a solution to the next order in $\k^2$? We
immediately see that the presence of $\RwR$ terms in the
Bianchi identity~\refs{Bianchi}, or equivalently in the boundary
conditions~\refs{bdryG}, which are non-zero for a general Calabi-Yau
space, mean that we can no longer set $G$ to zero. We can, however,
take a lead from the weakly coupled limit and try embedding the spin
connection in the gauge group. In the weakly coupled case, the
orbifold dimension is small and the low-energy theory is effectively
ten-dimensional. There is an analogous Bianchi identity for the
antisymmetric tensor, $dH\sim\FwFa+\FwFb-\RwR$. If we embed the spin
connection in one of the $E_8$ groups, say $F^{(1)}$, we can set
$\FwFa=\RwR$ and so set $H$ to zero. In the strongly
coupled limit, however, the same trick does not quite work. Because
the gravitational contribution to the Bianchi identity~\refs{Bianchi}
is distributed between the two hyperplanes, we do not find
$dG=0$. However, suppose we do embed the spin-connection in, say, the
$F^{(1)}$ group, and choose the internal background 
\be
 \FwFa=\RwR \qquad F^{(2)}_{AB}=0 \; ,
\ee
breaking the symmetry of one $E_8$ to $E_6$, while leaving the other
$E_8$ unbroken. We find there is then an equal and opposite
contribution to the Bianchi identity from the two hyperplanes,
$\frac{1}{2}\RwR$ from $M_1^{10}$ and $-\frac{1}{2}\RwR$ from
$M_2^{10}$. Witten's procedure is to solve the Bianchi identity to
this order, together with the equations of motion, with the additional
constraint that the solution continues to preserve $N=1$ supersymmetry
in four dimensions. To do this requires distorting the six-dimensional
manifold so that it is no longer a Calabi-Yau three-fold. 

The distortion turns out to be of the form 
\bea
 ds^2 &=& g^{(0)}_{IJ}dx^Idx^J + k_{IJ}dx^Idx^J \nn \\
      &=& g^{(0)}_{IJ}dx^Idx^J + \left(b\eta_{\m\n}dx^\m dx^\n 
          + h_{AB}dx^Adx^B + \g (dx^{11})^2 \right) 
      \; , \label{metric1}
\eea
where the corrections $b$, $h_{AB}$ and $\g$ depend on the Calabi-Yau 
coordinates and $x^{11}$, and only the $h_{a\bbar}$ terms are
non-zero. 

The result of solving for vanishing supersymmetry variation of the
gravitino to this order can be summarized as follows. Define the
quantities
\bea
 \b_A &=& \o^{BC}G_{ABC11} \nn \\
 \th_{AB} &=& w^{CD}G_{ABCD} \label{bta}\\
 \a &=& \o^{AB}\o^{CD}G_{ABCD} \nn \; ,
\eea
where we raise and lower all indices with the metric $\O_{AB}$.
The equation of motion and the Bianchi identity lead to the following
set of relations for the above quantities, as given by Witten~\cite{w},
\bea
 D_{\bar{a}}\b_{\bar{b}}-D_{\bar{b}}\b_{\bar{a}} &=& 0 \nn \\
 -\frac{i}{2}D^A\th_{A\bar{b}}+\frac{1}{4}D_{\bar{b}}\a
      -\frac{i}{2}D_{11}\b_{\bar{b}} &=& 0\nn\\
 D_{11}\b_{\bar{b}}-D^b\th_{b\bar{b}} &=& 0 \label{bta_eq}\\
 D_{11}\b_{\bar{a}}+\frac{i}{4}D_{\bar{a}}\a &=& 0\nn \\
 D^A\b_A &=& 0 \nn 
\eea
together with two more which come from contractions of the
$(dG)_{ABCD11}$ identity with $\o^{AB}\o^{AB}$ and $\o^{CD}$
respectively,
\be
  D_{11}\a + 4i\left(D^\abar\b_\abar - D^a\b_a\right) 
    = -\frac{1}{2\sqrt{2}\pi}\left(\frac{\k}{4\pi}\right)^{2/3}
         J \left\{\d (x^{11}) - \d (x^{11}-\pi\r)\right\}
    \label{BI1}
\ee
\bea
  D_{11}\theta_{a\abar} + \left(D_a\b_\abar - D_\abar\b_a\right) 
    &+& i\left(D^\bbar G_{\bbar a\abar 11} - D^b G_{ba\abar 11}\right) 
    \nn \\ && \quad = 
    -\frac{1}{2\sqrt{2}\pi}\left(\frac{\k}{4\pi}\right)^{2/3}
          J_{a\abar} \left\{\d (x^{11}) - \d (x^{11}-\pi\r)\right\}
    \; . \label{BI2}
\eea
The sources $J_{AB}$ and $J\equiv \o^{AB}J_{AB}$ are
given in terms of $J^{(i)}$ defined in equation~\refs{Jdef} as
follows. With the spin connection embedded in the gauge connection we
have 
\be
 J^{(1)}_{ABCD} = - J^{(2)}_{ABCD} \equiv J_{ABCD} 
      = \frac{1}{2}\left(\tr R^{(\O)}\wedge R^{(\O)} \right)_{ABCD} \; ,
\ee
where $R^{(\O)}_{AB}$ is the curvature of the zeroth-order metric on the 
Calabi-Yau space $\O_{AB}$. We then have
\be
  J_{AB} = J_{ABCD}\o^{CD} =
               3{\rm tr}R^{(\O)}_{[AB}R^{(\O)}_{CD]}\,\o^{CD} \; .
\ee
Note that it is only these last two equations for $G$~\refs{BI1}
and~\refs{BI2} which  receive contributions from the hyperplane terms
in the Bianchi identity~\refs{Bianchi}. 

Introducing a corrected spinor $\tilde{\eta}=e^{-\psi}\eta$ and
substituting into the supersymmetry variation~\refs{susy}, to order
$\k^{2/3}$ one finds that the variation vanishes if the following
conditions on the metric distortion are satisfied, 
\bea
 \sqrt{2}i\b_{\bar{a}} &=& 6\partial_{\bar{a}}b
   = -24\partial_{\bar{a}}\psi 
   = -3\partial_{\bar{a}}\g\label{susyeq1} \\
 \frac{1}{2\sqrt{2}}\a &=& 6\partial_{11}b = -24\partial_{11}\psi 
   \label{susyeq2} \\
 \partial_{\bar{a}}h_{b\bar{b}}-\partial_{\bar{b}}h_{b\bar{a}}&=&
   -\sqrt{2}\left( G_{b\bar{a}\bar{b}11}+\frac{i}{6}(
   \O_{b\bar{a}}\b_{\bar{b}}-\O_{b\bar{b}}\b_{\bar{a}})\right)
   \label{susyeq3} \\
 \O^{\bar{b}b}D_{\bar{b}}h_{b\bar{a}} &=&-\frac{\sqrt{2}i}{3}
   \b_{\bar{a}} \label{susyeq4} \\
 \partial_{11}h_{a\bar{b}} &=& -\frac{1}{\sqrt{2}}\left(i\th_{a\bar{b}}
   -\frac{1}{12}\a\O_{a\bar{b}}\right)\; . \label{susyeq5}
\eea
The full set of equations is compatible and thus a solution exists
which preserves the $N=1$ supersymmetry of the background. 

%%%%%%%%%%%%%%%%%%%%%%%%%%%%%%%%%%%%%%%%%%%%%%%%%%%%%%%%%%%%%%%%%%%%%%%%%

\section{Properties of Witten's solution}

\subsection{Explicit form of the solution}

It is possible to summarize Witten's solution in a more compact and
explicit form. 
To do this, we first dualize the four-form field strength $G$ to a
seven-form field $\cH$. This is possible because, for the configurations
we are considering, the $G\wedge G$ terms do not contribute to the
equation of motion for $G$. We define
\be
  \cH_{I_1\ldots I_7} = \ast G_{I_1\ldots I_7}
     = \frac{1}{4!}
       \e_{I_1\ldots I_8I_9I_{10}I_{11}}G^{I_8I_9I_{10}I_{11}} \; .
  \label{dual}
\ee
If we include the sources which live at the boundaries of the orbifold,
the Bianchi identity and equation of motion for $\cH$ read
\be
  d\cH = 0 \qquad D^{I_1}\cH_{I_1I_2\ldots I_6} = -\frac{1}{2\sqrt{2}\pi}
    \left(\frac{\k}{4\pi}\right)^{2/3} \left\{
        \ast J^{(1)}_{I_1\ldots I_6}\d(x^{11}) 
        + \ast J^{(2)}_{I_1\ldots I_6}\d(x^{11}-\pi\r) \right\} \; , 
\ee
where $\ast J^{(i)}$ is the dual of $J^{(i)}$ defined in
equation~\refs{Jdef}, so that 
\be
  \ast J^{(i)I_1\ldots I_6} = 
     \frac{1}{4}\e^{I_1\ldots I_6I_7I_8I_9I_{10}11}\left\{
          {\rm tr}F^{(i)}_{I_7I_8}F^{(i)}_{I_9I_{10}}
          - \frac{1}{2}{\rm tr}R_{I_7I_8}R_{I_9I_{10}} 
       \right\} \; . \label{Bsources}
\ee
The Bianchi identity is solved by writing 
$\cH_{I_1\ldots I_7}=7\partial_{[I_1}\cB_{I_2\ldots I_7]}$. We are then
also, as usual, free to choose a harmonic gauge where
$D^{I_1}\cB_{I_1\ldots I_6}=0$. The equation of motion then
reduces to 
\be
  \D_{11} \cB_{I_1\ldots I_6} = -\frac{1}{2\sqrt{2}\pi}
    \left(\frac{\k}{4\pi}\right)^{2/3} \left\{
        \ast J^{(1)}_{I_1\ldots I_6}\d(x^{11}) 
        + \ast J^{(2)}_{I_1\ldots I_6}\d(x^{11}-\pi\r) \right\} \; , 
\ee
where $\D_{11}$ is the eleven-dimensional Laplacian. 

Now let us specialize to the particular solution considered by
Witten. As we have already discussed, the zeroth-order solution is
simply the product of a Calabi-Yau three-fold and a $S^1/Z_2$
orbifold. As before, embedding the spin connection in the gauge
connection for $F^{(1)}_{IJ}$, we find that the non-zero components of
the sources~\refs{Bsources} are 
\be
  \ast J^{(1)}_{\m\n\r\s AB} = - \ast J^{(2)}_{\m\n\r\s AB} =
    \frac{1}{8}\e_{\m\n\r\s}\e_{ABCDEF}{\rm tr}R^{CD}R^{EF} 
    \; . \label{Wsource}
\ee
Thus we find that the only component of $B$ which is excited is a
(1,1)-form on the Calabi-Yau space, of the form
\be
  \cB_{\m\n\r\s a\bbar} = \e_{\m\n\r\s} \cB_{a\bbar} \; .
\ee
We can effectively ignore the external four-space and simply consider
a two-form potential in the internal seven-space. The corresponding
three-form field strength $\cH$ has the nonvanishing components
$\cH_{11AB}=\partial_{11}\cB_{AB}$ and
$\cH_{ABC}=3\partial_{[A}\cB_{CD]}$. Then, 
the relation~\refs{dual} between the four-form $G$ and $\cH$ turns into
\bea
 \cH_{11AB} &=& \frac{1}{24}\e_{ABCDEF}G^{CDEF} \nn\\
 \cH_{ABC} &=& \frac{1}{6}\e_{ABCDEF}G^{DEF11} \label{dual1}\; .
\eea
We can then express the fields $G_{\abar b\cbar 11}$,
$\q_{a\bbar}$, $\b_a$ and $\a$ defined in the previous section in terms
of $\cH$ as
\bea
 G_{\abar b\cbar 11} &=& i\cH_{\abar b\cbar}
    + \frac{1}{2}\left( \O_{\abar b}\cH_\cbar-\O_{\cbar b}\cH_\abar\right)
    \nn \\
 \q_{a\bar{b}} &=& - 2\cH_{11a\bar{b}} + \O_{a\bar{b}}\cH_{11} \nn \\
 \b_a &=& i\cH_a \label{dual2} \\
 \a &=& 4\cH_{11} \nn\; ,
\eea
where $\cH_A=\o^{BC}\cH_{ABC}$ and $\cH_{11}=\o^{BC}\cH_{11BC}$.

Since the sources only depend on the internal coordinates, the equation
of motion for the (1,1)--form $\cB$ reduces, at order $\k^{2/3}$, to 
\be
  \left(\D_X + D_{11}^2\right) \cB_{AB} = 
       - \frac{1}{16\sqrt{2}\pi}\left(\frac{\k}{4\pi}\right)^{2/3}
          \e_{ABCDEF}{\rm tr}R^{(\O)CD}R^{(\O)EF}
       \left\{ \d(x^{11}) - \d(x^{11}-\pi\r) \right\} \; . \label{Beq}
\ee
where $\D_X$ is the Laplacian on the Calabi-Yau space, while the gauge
condition on $B$ gives
\be
  D^A \cB_{AB} = 0 \; . \label{Bgauge}
\ee
These equations are completely equivalent to the set of equations of motion
and Bianchi identities given for $G$ in equations~\refs{bta_eq} above. 

As we have reviewed in the previous section, the correction to the metric
is fixed by requiring $N=1$ supersymmetry in four dimensions, which leads to 
the relations~\refs{susyeq1}--\refs{susyeq5}. In terms of the two-form
$\cB_{a\bbar}$, we find that these are equivalent to the simple relations
\bea
  h_{a\bbar} &=& \sqrt{2}i \left( \cB_{a\bbar} 
     - \frac{1}{3}\o_{a\bbar}\cB \right) + h'_{a\bbar} \nn \\
  b &=& \frac{\sqrt{2}}{6} \cB + b' \label{keq}\\
  \g &=& -\frac{\sqrt{2}}{3} \cB + \g' \nn  
\eea
and
\be
  \psi = -\frac{\sqrt{2}}{24} \cB + \psi' \; , 
\ee
where $\cB=\o^{AB}\cB_{AB}$. Here $h'_{a\bbar}$ is a zero mode of the
Laplacian on the Calabi-Yau space, $b'$ and $\psi'$ are constants and $\g'$
is an arbitrary function of $x^{11}$. Taking a trace, one finds the
useful relation
\be
 h = -\sqrt{2}\cB+h'\; ,\label{hB}
\ee
for $h=\O^{AB}h_{AB}$, where $h' =\O^{AB}h'_{AB}$.
To see the structure of this solution, we can solve the equation for
$\cB_{AB}$ explicitly in terms of eigenmodes of the Laplacian on the
Calabi-Yau space. We define 
\be
  \D_X \p^{i}_{a\bbar} = - \l_{i}^2 \p^i_{a\bbar} \; .
\ee
The eigenvalues are zero or negative since the Calabi-Yau three--fold is
compact. We will usually write the subset of eigenmodes with zero
eigenvalue as $\o^i_{a\bbar}$. The modes form an orthonormal set
normalized by
\be
  \int_X \p^{iAB}\p^j_{AB} = 6V \d^{ij} \; ,
\ee
where $V$ is the volume of the Calabi-Yau space. We can then decompose the
sources into eigenmodes, such that
\be
  - \frac{3}{2\sqrt{2}\pi}\left(\frac{\k}{4\pi}\right)^{2/3}
       \e_{ABCDEF}{\rm tr}R^{(\O)CD}R^{(\O)EF} =
       \sum_i \a_i \p^i_{AB} \; 
\ee
where the coefficients $\a_i$ are given by
\be
  \a_i = - \frac{1}{\sqrt{2}\pi V}\left(\frac{\k}{4\pi}\right)^{2/3}
       \int_X \p^i\wedge{\rm tr}R^{(\O)}\wedge R^{(\O)} \; . \label{alphai}
\ee
Matching the sources at $x^{11}=0$ and $x^{11}=\pi\r$, and recalling
that $\cB_{a\bbar}$  must be an even function under $Z_2$, we find an
explicit solution of the form
\be
  \cB_{a\bbar} = \sum_{\rm massive} \frac{\a_i\sinh \l_i(|x^{11}|-\p\r/2)}
          {24\l_i\sinh(\p\r\l_i/2)} \pi^i_{a\bbar}
    + \sum_{\rm massless} \frac{1}{24}\a_i\left(|x^{11}|-\p\r/2\right)
      \o^i_{a\bbar} + \cB '_{a\bbar} \; . \label{Bsol}
\ee
Here $\cB '_{a\bbar}$ is a general zero mode of the $\D_X+D_{11}^2$
operator. As such, it is pure gauge and will be set to zero from here
on.

We immediately notice some important properties of the
solution. First, the contribution from the massive modes decays as one
moves a distance of order $V^{1/6}$ away from the orbifold
planes. This is to be expected, since far from the planes the field
does not see the details of how the sources vary over the
Calabi-Yau space, but only their average properties characterized by their
zero-mode decomposition. Further, with these particular sources, which
are equal and opposite on the two planes, there is a gauge where the
correction $\cB_{a\bbar}$ is exactly zero at the middle of the orbifold
interval $x^{11}=\p\r/2$. If one would not use the standard embedding
as a starting point for the construction of the solution, the sources
would not be equal and opposite. Then, the $\a_i$ generally would not
be the same for the two sources and there would be no
point in the orbifold interval where the contributions of the massive
modes to $\cB_{a\bbar}$ could all be zero. Finally, one notes that in order 
to match the massless modes to the sources, the values of the massless
$\a_i$--coefficients must be the same for each source. This is simply
a realization of the condition, noted by Ho\v rava and Witten,
that the sources must be cohomologically zero. 

Let us now turn to discussing the structure of the metric correction
which, as we have seen, is essentially completely specified in terms of
$\cB_{a\bbar}$. The only remaining freedom are the functions $h'$,
$b'$ and  $\g'$. However, these simply reflect an ambiguity in how one
defines the zeroth-order solution. There is always a freedom in the
first-order solution corresponding to zero-mode deformations of the
original space, such as rescaling the volume of the Calabi-Yau or the
length of the orbifold interval. However, these corrections are most
naturally included in the definition of the zeroth-order solution, and
so set to zero. To see this explicitly, consider defining a new
zeroth-order metric by
\be
  ds^2 = (1+b')\eta_{\m\n}dx^\m dx^\n + (\O_{AB}+h'_{AB})dx^Adx^B
            + (1+\g')\left(dx^{11}\right)^2 \; . \label{g0prime}
\ee
It is easy to see that this has exactly the same form as the original
zeroth-order metric~\refs{metric0}. First we note that, since $b'$ is a
constant, it can be  absorbed  by rescaling the $x^\m$
coordinates. Since we are free to refine the $x^{11}$ coordinate by
$x^{11}\rightarrow f(x^{11})$, we can similarly remove the arbitrary
function $\g'(x^{11})$. Finally, since $h'_{AB}$ is a zero mode of the
Calabi-Yau Laplacian, the new metric $\O'=\O+h'$ still describes a
Calabi-Yau manifold. We might as well redefine the zeroth-order metric
by equation~\refs{g0prime} and then set $b'=\g'=h'_{AB}=0$ in the
first-order solution. 

Of course, other definitions are possible. One could, for instance,
keep part of $h'_{AB}$ in the zeroth-order metric and part in the
first-order correction. However, there is one sense in which the above
prescription is natural. We have noted that in the gauge $\cB '_{AB}=0$,
the two-form $\cB_{AB}$  vanishes at the midpoint of the orbifold
interval. Further, with the prescription $b'=\g'=h'_{AB}=0$, we find
that the metric correction also vanishes at this point. That is to
say, at $x^{11}=\p\r/2$ the full first-order metric is simply that of
a Calabi-Yau cross an orbifold interval. Our prescription corresponds
to setting the zeroth-order solution equal to the metric at this
midpoint of the orbifold. We note that this property is special to
Witten's particular solution. Generically, when the sources on the two
hyperplanes are not equal and opposite, then there is no point in the
orbifold interval where the first-order metric reduces to that of
simply a Calabi-Yau space cross an orbifold interval. 

In summary, we can choose a gauge for $\cB_{AB}$ and a prescription for
the splitting of the metric into zeroth- and first-order pieces such that
Witten's solution is completely specified by the solution for $\cB_{AB}$
given in equations~\refs{alphai} and~\refs{Bsol}. The metric distortion
is related to $\cB_{AB}$ by the expressions~\refs{keq} and the various
components of $G_{IJKL}$ can be obtained from the eqs.~\refs{dual1}
and \refs{dual2}. These corrections all vanish at the
midpoint of the orbifold interval. Furthermore, we see from the form of the
solution for $\cB_{AB}$~\refs{Bsol}, that the average of the corrections
over the orbifold interval all vanish. That is 
\be
  \orbav{\cB_{AB}}=\orbav{k_{IJ}}=0 \label{vanish}
\ee
where we define the orbifold average by 
\be
 \orbav{F} = \frac{1}{\pi\r}\int_{0}^{\pi\r}dx^{11}F \; .
\ee
This property is peculiar to Witten's solution. More general sources
lead to first-order corrections which do not average to zero.

\subsection{The metric as a solution of Einstein's equations}

Let us now prove that Witten's background solves the equations of motion
to linear order. In doing so, we will find the solution matches to a
set of source terms, localized on the orbifold hyperplanes, which
exactly match the sources which would arise from the ten-dimensional
gauge fields and putative $R^2$ terms. This gives further evidence for
Gauss-Bonnet curvature terms in the eleven-dimensional action,
localized on the orbifold hyperplanes. 

We recall that the relations~\refs{keq} were
derived by requiring the solution had $N=1$ supersymmetry in four
dimensions working to order $\k^{2/3}$. Let us, for the moment, work
away from the orbifold hyperplanes. For consistency, the metric should
satisfy the Einstein equations~\refs{geom} to the same order in
$\k^2$. Since $G$ is zero to first order, the first non-zero
contribution to the stress-energy, which is quadratic in $G$, is at
second-order, proportional to $\k^{4/3}$. Thus we see that, to first
order in $\k^{2/3}$, the metric must satisfy the free Einstein
equations, namely 
\be
 R_{IJ} = 0 \; .
\ee
The zeroth-order metric $g^{(0)}_{IJ}$ is Ricci flat. Thus we are left
with the familiar condition that the first-order perturbation $k_{IJ}$
satisfies the linearized Einstein equations~\cite{wald} 
\be
 D^2 k_{IJ} + D_ID_J k - D^KD_I k_{JK} - D^KD_J k_{IK} = 0 \label{linEinstein}
\ee
where $D_I$ is the covariant derivative with respect to the zeroth-order 
metric and all index contractions are with $g^{(0)}_{IJ}$, so that,
for instance, $k=g^{(0)IJ}k_{IJ}$. Thus, for consistency of Witten's
solution, the relations~\refs{keq} on $b$, $\g$ and $h_{AB}$ derived
from the supersymmetry condition, together with the equation of
motion~\refs{Beq} and gauge condition~\refs{Bgauge} for $\cB_{AB}$, must
imply that the linearized Einstein equations~\refs{linEinstein} are
satisfied.  

To see this, we first note that the gauge condition for
$\cB_{AB}$~\refs{Bgauge} implies a gauge condition for $k_{IJ}$
\be
  D^I\left(k_{IJ}-\frac{1}{2}g^{(0)}_{IJ}k\right) = 0 \; .
  \label{gauge_cond}
\ee
We can use this to simplify the linearized Einstein
equation~\refs{linEinstein}. Together with the fact that the
zeroth-order metric is Ricci-flat, it means that the equation can be
rewritten as  
\be
  D^2 k_{IJ} + 2 R^{(0)}_{IKJL}k^{KL} = 0 \; . \label{k_eq}
\ee
Substituting from the expressions~\refs{keq} for $k_{IJ}$ in terms of
$\cB_{a\bbar}$, we find that the $b$ and $\g$ equations reduce to 
\be
  \left(\D_X+D_{11}^2\right) \cB = 0 \; ,
\ee
while the $h_{a\bbar}$ equation implies that 
\be
  \left(\D_X+D_{11}^2\right) \cB_{a\bbar} = 0 \; .
\ee
But from the equation of motion for $\cB_{a\bbar}$, we know that both of
these conditions are satisfied, at least away from the fixed orbifold
planes. Thus, the first-order solution for $k_{IJ}$,
satisfying the supersymmetry conditions~\refs{keq}, is also the
first-order solution to the eleven-dimensional equations of motion.  

What if, however, we include the source terms which appear in the
equation of motion for $\cB_{a\bbar}$? Calculating the linearized
components of the Einstein tensor $G_{IJ}=R_{IJ}-\frac{1}{2}g_{IJ}R$,
one finds the only non-zero components are
\bea
  G_{AB} &=& -\frac{1}{8\pi}\left(\frac{\k}{4\pi}\right)^{2/3}\left\{
     2 \tr R^{(\O)}_{AC}{{R^{(\O)}}_B}^C 
        - \frac{1}{2}g_{AB}\tr R^{(\O)}_{CD}R^{(\O)CD} \right\}
     \left(\d(x^{11}) - \d(x^{11}-\pi\r) \right) \nn \\
  G_{\m\n} &=& -\frac{1}{8\pi}\left(\frac{\k}{4\pi}\right)^{2/3}\left\{
     - \frac{1}{2}\eta_{\m\n}\tr R^{(\O)}_{CD}R^{(\O)CD} \right\} 
     \left(\d(x^{11}) - \d(x^{11}-\pi\r) \right) \; .
\eea
We recall that embedding the gauge connection in the spin connection
implies that $\tr R^{(\O)}_{AC}R^{(\O)C}_B=
\tr F^{(1)}_{AC}F^{(1)C}_B$. Furthermore, the zeroth order
curvature $R^{(\O)}_{AB}$ is Ricci flat. One then sees that these
sources match exactly to the sources that would arise from boundary
gauge and curvature Gauss-Bonnet terms of the form given in
equations~\refs{LYM} and~\refs{SR2} above. (Strictly, we only
constrain the Riemann curvature terms in the Gauss-Bonnet
combination, since the zeroth-order Ricci curvature vanishes.) This is
the second piece of evidence for $R^2$ terms at order $\k^{2/3}$ in
the effective action of M theory on an $S^1/Z_2$ orbifold.

\subsection{Setup for the dimensional reduction}

In this final subsection, we will prepare all the relevant information
which we are going to need for the computation of the effective
four-dimensional action. We will perform this computation up to linear
terms in the distortion of the background; that is, generally, up to terms
of order $\k^{2/3}$ from the bulk and of order $\k^{4/3}$ from the boundary.
Given that the eleven-dimensional theory is defined up to the order $\k^{2/3}$
only, it is difficult to reliably determine terms quadratic in the distortion
and we will not attempt to do so in this paper.

One set of fields which necessarily survive as massless fields in a
four-dimensional action are the moduli of the background solution. To
calculate the kinetic terms for these moduli, as well as their
interactions with other four-dimensional fields, we need to understand
how they enter the background solution. To zeroth-order, the moduli
simply correspond to the moduli of the Calabi-Yau manifold together
with a modulus describing a rescaling of the size of the orbifold. We
will generally concentrate in what follows on the generic
four-dimensional fields, which are independent of the particular form
of the Calabi-Yau manifold. For the zeroth-order metric, there are only
two such moduli controlling the overall size of the Calabi-Yau space and the
length of the orbifold interval. We write 
\be
  ds^2 = \eta_{\m\n} dx^\m dx^\n 
     + e^{2a}\O_{AB} dx^Adx^B + e^{2c} (dx^{11})^2 \; , \label{gmoduli}
\ee
so that the Calabi-Yau volume is now $e^{6a}V$ and the length of the orbifold
interval is $e^c\pi\r$. Corresponding moduli are present in the
first-order background, but they enter the metric in a much more
complicated way, since the correction does not depend on the
form of the zeroth-order solution in a simple way. Nonetheless, we can
derive the dependence on $a$ and $c$ using the explicit solution for
$\cB_{a\bbar}$~\refs{Bsol}. Since we take $h'=b'=\g'=0$, the metric
correction can be computed directly from $\cB_{a\bbar}$ using the
relations~\refs{keq}. We find the explicit expression 
\be
  \cB_{a\bbar} = e^{c-4a}
    \sum_{\rm massive} \frac{\a_i\sinh e^{c-a}\l_i(|x^{11}|-\p\r/2)}
          {24e^{c-a}\l_i\sinh(\p\r e^{c-a}\l_i/2)} \pi^i_{a\bbar}
    + e^{c-4a}\sum_{\rm massless}\frac{1}{24}
      \a_i\left(|x^{11}|-\p\r/2\right)\o^i_{a\bbar} \; . \label{Bsolmod}
\ee

As we have already noted, we would like to calculate the effective action to
linear order in the background distortion. Further, reducing to four
dimensions means integrating over the Calabi-Yau space and, for terms coming
from the bulk, also integrating over the orbifold interval. Thus we are
going to need averages of the correction over the Calabi-Yau space. Let us
define such an average by 
\be
 \cyav{F} = \frac{1}{V}\int{}d^6x\sqrt{\O}F \; . 
\ee
Then, recalling that $\o_{AB}$ is a zero mode of the Calabi-Yau
Laplacian, from the orthogonality of the eigenmodes we find that
\be
 \cyav{\cB} = \cyav{\o^{AB}\cB_{AB}}
     = \frac{1}{4}e^{c-4a}\a_0 \left(|x^{11}|-\p\r/2\right)
 \label{B_av}
\ee
where $\a_0$ is the coefficient of the $\o_{AB}$ zero mode in the
expansion of the source. That is 
\be
 \a_0 =  - \frac{1}{\sqrt{2}\pi V}\left(\frac{\k}{4\pi}\right)^{2/3}
       \int_X \o\wedge{\rm tr}R^{(\O)}\wedge R^{(\O)} \; .
 \label{a_sol}
\ee
Note that $\a_0$ also equals the Calabi--Yau average of the quantity $\a$
defined in eq.~\refs{bta}; that is,
\be
 \a_0 = \cyav{\a}\; .
\ee
We see that the moduli enter in exactly the combination of scales depending
on the ratio of the length of the orbifold interval to the volume of
the Calabi-Yau to the power 2/3, which appeared in the expansion
parameter $\e$ defined in the introduction. This is confirming that
Witten's solution is an expansion in $\e$.
Other relevant Calabi--Yau averages are those for $h$, $b$ and $\g$,
which can be directly obtained from eq.~\refs{B_av} and the
relations~\refs{keq}, \refs{hB}. One finds that
\bea
 \cyav{b} &=& \frac{\sqrt{2}}{24}e^{c-4a}\a_0(|x^{11}|-\pi\r /2) \nn \\
 \cyav{\g} &=& -\frac{\sqrt{2}}{12}e^{c-4a}\a_0(|x^{11}|-\pi\r /2)
  \label{CY_av1} \\
 \cyav{h} &=& -\frac{1}{2\sqrt{2}}e^{c-4a}\a_0(|x^{11}|-\pi\r /2) \nn\; .
\eea

The reason for the simple moduli dependence of the above averages is that,
by orthogonality, we have projected out the K\"ahler form term in the
zero mode part of the solution~\refs{Bsol}. The fact that the heavy
modes drop out is actually a general property of the effective action
up to linear terms in the distortion. This can be seen as
follows. Consider inserting the distorted background, which, since we
have set $h'=b'=\g '=0$ in eq.~\refs{keq}, can be
completely expressed in terms of $\cB_{AB}$, into the action and expand up to
linear terms in $\cB_{AB}$. Then the action will consist of zeroth order,
$\cB_{AB}$--independent pieces and terms of the form $L^{AB}\cB_{AB}$ where
$L^{AB}$ is some expression which depends on the zeroth order background only.
Consequently, $L^{AB}$ can be expanded in terms of the zero modes $\o^i_{AB}$
on the Calabi--Yau space as $L_{AB}=\sum_i L_i\o^i_{AB}$, where the
$L_i$ do not depend on Calabi--Yau coordinates. After integrating over
the Calabi--Yau space we have, therefore,
$\cyav{L^{AB}\cB_{AB}}=\frac{V}{4}e^{c-4a}\sum_iL_i\a_i(|x^{11}|-\pi\r /2)$;
that is, we project onto the zero mode part of $\cB_{AB}$ in
eq.~\refs{Bsolmod}. In addition, if the term under consideration results from
the bulk action, we have to perform another average over the orbifold
direction so that $\av{L^{AB}\cB_{AB}}=0$, as long as the $L_i$ are
$x^{11}$ independent. Though the latter condition will be ordinarily
fulfilled for a reduction to four dimensions (since all the fields just
depend on $x^\m$), we will find that the Bianchi identity forces us to
introduce $x^{11}$--dependent background fields. In those cases, a nonzero
result (from background zero mode pieces) remains after performing the
orbifold average. As we have seen previously, the vanishing of the
orbifold average is related to the specific choice of the solution we have
made by setting $h'=b'=\g ' =0$ in the relations~\refs{keq}. For this choice,
the orbifold average of the metric correction $k_{IJ}$ vanishes; that is,
$\orbav{k_{IJ}}=0$. Another way of characterizing this choice is therefore the
following. The moduli $a$ and $c$ are defined in such a way that the volume
of the original Calabi-Yau space $e^{6a}V$ and the length of the original
orbifold interval $e^c\pi\r$ equal the corresponding quantities for
the distorted background averaged over the orbifold and the Calabi-Yau
space respectively. 

To summarize, we have seen that, to linear order in the distortion,
the heavy modes in the expansion for $\cB_{AB}$, eq.~\refs{Bsolmod} do not
contribute to the effective action. In addition, for the part of the
effective action arising from the bulk, the zero mode
pieces also vanish for the definition of moduli we have adopted. We expect,
however, corrections from the boundary actions, since no orbifold
average has to be performed in this case. The second source of corrections
are the $x^{11}$--dependent background fields mentioned above. The fact that
both types of corrections depend on the massless piece of the
solution~\refs{Bsolmod} only, simplifies some of the calculations for the
effective action considerably. Therefore, it is useful to analyze which of
the quantities describing the full background really depend on the massless
part of $\cB_{AB}$. Since the metric corrections $k_{IJ}$ are given directly
in terms of $\cB_{AB}$, they depend on massless as well as massive modes.
For the form field we note that, while $\cH_{11AB}$ 
contains all modes, $\cH_{ABC}$ depends on heavy modes only. Therefore,
$G_{ABCD}$, $\q_{AB}$, $\a$ consist of the full spectrum whereas
$G_{ABC11}$, $\b_A$ contain massive Calabi--Yau modes only. This
observation allows us to neglect $G_{ABC11}$ and $\b_A$ wherever necessary.

%%%%%%%%%%%%%%%%%%%%%%%%%%%%%%%%%%%%%%%%%%%%%%%%%%%%%%%%%%%%%%%%%%%%%%%%%%%%%%

\section{The $D=4$ effective action}

We now turn to deriving the four-dimensional effective action by
reducing the general eleven-dimensional action~\refs{LYM} on Witten's
background solution. By construction, the low-energy massless
excitations fall into four-dimensional $N=1$ chiral or vector
supermultiplets. The low-energy theory is then characterized by
specifying three types of functions. The K\"ahler potential $K$
describes the structure of the kinetic energy terms for the chiral
fields while the holomorphic superpotential $W$ encodes the potential
for the chiral fields. The holomorphic gauge kinetic functions $f$ represent
the coupling of the chiral multiplets to the $E_6$ and $E_8$ gauge fields. For
a set of chiral fields $Y^\imath$, the relevant terms in the low-energy
action are 
\bea
  S &=& - \frac{1}{16\pi G_N}\int_{M^4} \sqrt{-g}\left[ R+ 
          2K_{\imath\bar{\jmath}}\partial_\mu Y^\imath \partial^\mu
          \bar{Y}^{\bar{\jmath}}
          +2e^K\left( K^{\imath\bar{\jmath}}D_\imath W\overline{D_\jmath W}
          -3|W|^2\right)+\mbox{D-terms}\right]\nn \\
    &&  -\frac{1}{16\pi\a_{\rm GUT}}\int_{M^4}\sqrt{-g}\left[
          {\rm Re}f(Y^\imath){\rm tr}F^2 +{\rm Im}f(Y^\imath){\rm tr}F\Fd
         \right] \; , \label{SUGRA}
\eea
Here $K_{\imath\bar{\jmath}}=\frac{\partial^2 K}{\partial Y^\imath
\partial\bar{Y}^{\bar{\jmath}}}$ is the K\"ahler metric and
$D_\imath W=\partial_\imath W+\frac{\partial K}{\partial Y^\imath}W$ is
the K\"ahler covariant derivative acting on the superpotential. The dual
field strength $\Fd_{\m\n}$ is defined as
$\Fd_{\m\n}=\frac{1}{2}\e_{\m\n\s\r}F^{\s\r}$. Note that in the above
expression all fields $Y^\imath$ are chosen to be dimensionless.

In this section, we will derive these functions, first to
zeroth-order and then include the $\k^{2/3}$ corrections. Finally, we will
deduce some of the terms arising from gauge fields at order $\k^{4/3}$.
In each case, one must first identify the zero mode excitations for the given
background and then calculate the effective action. We will
concentrate on the generic modes, which are independent of the
particular form of the Calabi-Yau. 

\subsection{Zeroth order action}

To zeroth-order the background solution completely factorizes into a
Calabi-Yau space times an orbifold. The relative size of the two
spaces does not enter the solution. For this reason, the effective action is
identical to the weakly coupled limit, which corresponds to taking
the orbifold size to be small. The derivation is consequently well
known. We will, nevertheless, repeat the essential steps for
completeness and to set the notation.

We start by giving the zeroth-order background and identifying the
corresponding zero modes. The metric has the form
\be
  ds^2 = \bar{g}_{\m\n}dx^\m dx^\n 
    + e^{2a}\O_{AB} dx^Adx^B
    + e^{2c}(dx^{11})^2 \; . \label{g0mod}
\ee
Here $\O_{AB}$ is the metric of the background Calabi-Yau space, while
$\bar{g}_{\m\n}$ is the metric in the external four-dimensional
space. The universal moduli are $a$, measuring the volume of the
Calabi-Yau, and $c$, measuring the size of the orbifold interval.

The components $G_{ABCD}$, $G_{ABC11}$ of Witten's background can be
neglected to this order, while the gauge fields
do not enter until we go to order $\k^{2/3}$. The remaining low-energy
bosonic fields come from components of $G$ which survive
the $Z_2$ orbifold projection. As there are no harmonic one-forms on
a generic Calabi-Yau, the only relevant components are 
$G_{\m AB11}$ and $G_{\m\n\r 11}$. Since the sources vanish to zeroth
order, the Bianchi identity~\refs{Bianchi} implies that 
$\partial_{[\m}G_{\n]AB11}=\partial_{[\m}G_{\n\r\s]11}=0$, so that we
can write the field strengths in terms of the three-form potential as
\be
 G_{\m AB11} = 6\partial_\m C^{(0)}_{AB11} \; ,\quad
 G_{\m\n\r 11} = 18\partial_{[\m}C^{(0)}_{\n\r]11} \; .\quad
 \label{form0_4}
\ee
The equations of motion then imply that the zero modes for $C^{(0)}_{AB11}$
correspond to harmonic two-forms in the Calabi-Yau. Since the K\"ahler
form is harmonic, we always have the modes 
\be
 C^{(0)}_{AB11} = \frac{1}{6}\c \o_{AB}\; ,\quad
 C^{(0)}_{\m\n 11} = \frac{1}{6}B_{\mu\nu} \; . \label{C0mod}
\ee
Thus the universal low-energy moduli are the scalars $a$, $c$, $\c$
and the two--form $B_{\m\n}$. 

With the given expression for the metric and the components of $G_{IJKL}$,
the eleven-dimensional supergravity action~\refs{LSG} truncates to
\be
 S^{(0)} = \frac{\pi\r V}{\k^2}\int_{M^4} \sqrt{-g}\left[
                    -R-18\partial_\m a\partial^\m a -
                    \frac{3}{2}\partial_\m\hat{c}\partial^\m\hat{c} -
                    3e^{-2\hat{c}}\partial_\m\c\partial^\m\c
                    -\frac{1}{6}e^{12a}H_{\m\n\r}H^{\m\n\r}
                    \right]\; ,
 \label{action4}
\ee
where we have already performed the Weyl rotation
\be
 \bar{g}_{\m\n} = e^{-6a-c}g_{\m\n}
\ee
to the Einstein frame metric $g_{\m\n}$. The modulus $\hat{c}$ is defined by
\be
 \hat{c}=c+2a\; ,
\ee
while $H_{\m\n\r}=3\partial_{[\m} B_{\n\r]}$. By adding the term
$\left(\pi\r V/6\k^2\right)\int d^4x\sqrt{-g}\s\e^{\m\n\r\s}\partial_\m 
H_{\n\r\s}$
to the action~\refs{action4}, we can dualize the two-form $B_{\m\n}$ to
a scalar field $\s$. From the equation of motion for $H_{\m\n\r}$, one then
finds
\be
 H_{\m\n\r} = e^{-12a}{\e_{\m\n\r}}^\s\partial_\s\s\; ,
\ee
which leads to
\be
 S^{(0)} = \frac{\pi\r V}{\k^2}\int_{M^4}\sqrt{-g}\left[
                    -R-18\partial_\m a \partial^\m a -
                    \frac{3}{2}\partial_\m\hat{c}\partial^\m\hat{c} -
                    3e^{-2\hat{c}}\partial_\m\c\partial^\m\c
                     -e^{-12a}\partial_\m\s\partial^\m\s
                    \right]\; .
 \label{act4_dual}
\ee
We note the appearance of the expression for the four-dimensional
gravitational coupling first given in equation~\refs{couplings} above. 
By comparison with eq.~\refs{SUGRA}, it is easy to see that this action can
be derived as the bosonic part of an $N=1$ supergravity action with the
familiar K\"ahler potential
\be
 K = -\ln (S+\bar{S})-3\ln (T+\bar{T})
 \label{kahler}
\ee
if the field identification
\be
 S=e^{6a}+i\sqrt{2}\s\; ,\quad T = e^{\hat{c}}+i\sqrt{2}\c \label{ST}
\ee
is made. 

\subsection{Distorted background and zero modes }

In this section, we will derive the complete solution which we are going to
use for the calculation of the order $\k^{2/3}$ and $\k^{4/3}$ corrections.
Several new features with respect to the zeroth-order calculation of
the previous subsection have to be considered. Clearly, the corrections
to the Calabi--Yau background metric have to be taken into account.
Similar corrections can be expected for the zero modes coming from the
three--form. At the order $\k^{2/3}$, we also have to consider the gauge
fields which give rise to four--dimensional gauge fields and gauge matter.
As we will see, these additional nonzero components of the gauge fields
complicate the picture even further. Via the nontrivial Bianchi identity
and the source terms in the Einstein equation, these gauge fields switch
on a new background configuration of both the four-form field strength and
the metric, which will play a crucial r\^ole in deriving 
the effective action.

Let us start by considering the zero modes from the gauge fields; that is,
the four--dimensional gauge fields $A^{(1)}_{\m}$, $A^{(2)}_{\m}$ with 
field strengths $F^{(1)}_{\m\n}$, $F^{(2)}_{\m\n}$ and the gauge--matter
fields. For the latter, we will concentrate on the generic zero mode, so
that our Ansatz for the internal part of the gauge field in the
zeroth-order background Calabi-Yau is given by
\be
 A^{(1)}_b=\bar{A}_b+{w_b}^cT_{cp}C^p\; . \label{gauge_matter}
\ee
Here $\bar{A}_b$ is the background field resulting from the standard
embedding (it equals the spin connection), $T_{cp}$ are the broken
$E_8$ generators transforming in the $({\bf 3},{\bf 27})$
representation of $SU(3)\times E_6$ and $C^p$ is the
generic matter field transforming in the ${\bf 27}$ representation of
$E_6$. The generators are normalized so that 
$\mbox{tr}(T_{cp}\bar{T}^{dq})=\d_c^d\d_p^q$. Another useful relation
is $\mbox{tr}(T_{ap}T_{bq}T_{cr}) =\e_{abc}d_{pqr}$,
where $d_{pqr}$ is the tensor which projects out the singlet in
${\bf 27}^3$. Since the gauge fields appear at order $\k^{2/3}$,
one does not need to worry about corrections to eq.~\refs{gauge_matter},
due to the distortion of the background metric, for a calculation of
the effective action to this order. We will argue
below, however, that for terms involving gauge fields we can reliably
calculate the effective action to the order $\k^{4/3}$. Clearly, for
those terms, order $\k^{2/3}$ corrections to eq.~\refs{gauge_matter}
are relevant and, in general, such corrections can be expected since
the $\k^{2/3}$ metric distortion modifies the equation of motion for
$A^{(1)}_b$. We can account for this modification by replacing the K\"ahler
form $\o_{a\bar{b}}$ in eq.~\refs{gauge_matter} with a corrected $(1,1)$ form
$\o_{a\bar{b}}+\hat{\o}_{a\bar{b}}$, which is harmonic with respect to
the distorted metric. Note that, since we are dealing with the gauge field
at the boundary $x^{11}=0$, this distorted metric corresponds to the
deformed Calabi--Yau space at $x^{11}=0$. Therefore, if we expand
the correction $\hat{\o}_{a\bar{b}}$ as
$\hat{\o}_{a\bar{b}}=\sum_{\rm massive}\l_i\pi^i_{a\bar{b}}+
\sum_{\rm massless}\l_i\o^i_{a\bar{b}}$ into harmonics on the Calabi--Yau
space, the expansion coefficients $\l_i$ are $x^{11}$--independent.
Consequently, the massless terms can be absorbed by a redefinition of the
moduli. As a result, the Ansatz~\refs{gauge_matter} only receives
corrections corresponding to heavy Calabi--Yau modes. 
The massive coefficients $\l_i$ in this expansion, though not
needed in the following, can be computed explicitly by using the
solution~\refs{Bsolmod} for Witten's background. 
These, however, can be neglected since they are orthogonal to the
zeroth-order zero-mode given in~\refs{gauge_matter} and so, after
integrating over the Calabi-Yau, their linear-order contribution
vanishes. In the following, we can, therefore, simply work with
eq.~\refs{gauge_matter}. The nonvanishing components of the field
strength are then given by
\bea
 F^{(1)}_{ab} &=& {\o_a}^c{\o_b}^d[T_{cp},T_{dq}]C^pC^q \nn \\
 F^{(1)}_{a\bar{b}} &=& \bar{F}_{a\bar{b}}+
                  {\o_a}^c\o_{\bar{b}d}[T_{cp},\bar{T}^{dq}]C^p\bar{C}_q 
 \label{Fs} \\
 F^{(1)}_{\m b} &=& {\o_b}^cT_{cp}(D_\m C)^p\nn\; .
\eea
Another useful relation for the background $\bar{F}_{AB}$ is
\be
 \cyav{\mbox{tr}\bar{F}_{AB}\bar{F}^{AB}}=\cyav{\mbox{tr}R^{(\O )}_{AB}
 R^{(\O )AB}} = 2\sqrt{2}\pi (4\pi /\k )^{2/3}\a_0\; .\label{FR}
\ee

We now turn to a discussion of the corrections to the metric. Generally,
we split the metric into three pieces
\be
 g_{IJ} = g_{IJ}^{(0)}+g_{IJ}^{(1)}+g_{IJ}^{(B)}\; \label{mgen}
\ee
Here, $g_{IJ}^{(0)}$ and $g_{IJ}^{(1)}$ are the familiar zeroth and
first order pieces which we have discussed previously. Introducing the
generic moduli $a$ and $c$ in eq.~\refs{metric1}, they read
\bea
 g^{(0)}_{\m\n} = \bar{g}_{\m\n}\; ,&g^{(0)}_{AB}=e^{2a}\O_{AB}\; ,&
 g^{(0)}_{11,11} = e^{2c} \nn \\
 g^{(1)}_{\m\n} = b(a,c)\bar{g}_{\m\n}\; ,&g^{(1)}_{AB} = e^{2a}h_{AB}(a,c)
 \; ,& g^{(1)}_{11,11} = e^{2c}\g (a,c)\; .
 \label{metric3}
\eea
As we have explained in section 3, the distortion $b$, $h_{AB}$ and $\g$
have an implicit dependence on $a$, $c$ which we have indicated
in eq.~\refs{metric3}. Its explicit form can be obtained from
the previous results by expressing the distortion in terms of the
two--form $\cB_{AB}$ via eq.~\refs{keq}, and using the solution~\refs{Bsolmod}
for $\cB_{AB}$. This solution contains the full dependence on the generic
moduli $a$ and $c$. It is with these replacements understood,
that eq.~\refs{metric3} should be used for the calculation
of the effective action. For the effective action, we will consider
corrections linear in the background distortion. In such a situation,
we have shown that the heavy Calabi--Yau modes in the solution for $\cB_{AB}$
do not contribute to the effective action, which simplifies the
problem considerably. Correspondingly, from eq.~\refs{Bsolmod}, the moduli
dependence of the massless part of $b$, $h_{AB}$ and $\g$ is simply a
scaling with $e^{c-4a}$.

We have not yet discussed the last piece, $g_{IJ}^{(B)}$, in the
metric~\refs{mgen}. It arises through a mechanism analogous to that
which leads to the Chern-Simons and Green-Schwarz terms in the
reduction from eleven to ten dimensions discussed in section 2.2. As explained
there, this piece is explicitly $x^{11}$--dependent and originates
from the gauge field source terms in the Einstein equation~\refs{geom}.
The physical picture is that small fluctuation of the observable gauge and
gauge matter fields cause boundary source terms which force the metric
(and the four--form as we will see below) to interpolate between those
sources. This can be viewed as a kind of back--reaction, where every
fluctuation of a low energy gauge field causes a small distortion of
the background on which the reduction is carries out. Clearly, to
arrive at a sensible purely four--dimensional effective action, this
back--reaction has to be taken into account. To determine $g_{IJ}^{(B)}$,
we write the Einstein equation~\refs{geom} to the order $\k^{2/3}$
in the form
\bea
 R_{IJ} &+& \frac{1}{6}\left(G_{IKLM}{G_J}^{KLM}
       -\frac{1}{12}g_{IJ}G^2\right) \nn \\
   &&= -\frac{1}{2\pi}(\k /4\pi )^{2/3}\left(\d (x^{11})S^{(1)}_{IJ}+
          \d (x^{11}-\pi\r ) S^{(2)}_{IJ}\right)
 \label{geom1}
\eea
where the sources $S^{(i)}_{IJ}$ are given by
\bea
 S^{(i)}_{\bar{I}\bar{J}} &=& (g_{11,11})^{-1/2}\left(\mbox{tr}
     F_{\bar{I}\bar{K}}^{(i)}F_{\bar{I}}^{(i)\bar{K}}-\frac{1}{12}
     g_{\bar{I}\bar{J}}\mbox{tr}(F^{(i)})^2\right) \label{Sdef}\\
 S_{11,11}^{(i)} &=& \frac{1}{6}\sqrt{g_{11,11}}\mbox{tr}(F^{(i)})^2\; .
\eea
We are now going to solve this equation in a linearized form for the
correction $g^{(B)}_{IJ}$ to the metric, writing 
$g_{IJ}= g^{(0)}_{IJ}+g^{(B)}_{IJ}$. In order for the zero mode
fields in $g^{(0)}_{IJ}$ to be well defined, we require in addition that
$\orbav{g^{(B)}_{IJ}}=0$. To zeroth-order we neglect the source terms
and the zeroth-order metric satisfies the left-hand-side
of~\refs{geom1}. Expanding to first order, we then get an equation for
the contribution $g^{(B)}_{IJ}$ which now contains the extra
source term. Furthermore, since the source terms arise from slowly varying
low energy fields, we can neglect derivatives $D_\m$ in this equation.
Calabi--Yau derivatives, on the other hand, cannot be neglected,
since the radius of the Calabi--Yau space is smaller than the orbifold
length. We will, however, see later on that the Calabi--Yau part of the
source terms and, consequently, $g^{(B)}_{AB}$ is proportional to $\O_{AB}$.
Therefore, the Calabi--Yau part of the linearized Einstein equation vanishes
for this solution. Having said all this, the boundary value problem we
have to solve can be formulated in the ``downstairs'' picture as~\cite{llo}
\bea
 D_{11}^2g^{(B)}_{IJ} &=&\frac{1}{2\pi^2\r}(\k /4\pi )^{2/3}
                         \left( S^{(1)}_{IJ}+S^{(2)}_{IJ}\right)\nn \\
 \left. D_{11}g^{(B)}_{IJ}\right|_{x^{11}=0} &=& -\frac{1}{2\pi}
        (\k /4\pi )^{2/3}S^{(1)}_{IJ} \\
 \left. D_{11}g^{(B)}_{IJ}\right|_{x^{11}=\pi\r} &=& \frac{1}{2\pi}
        (\k /4\pi )^{2/3}S^{(2)}_{IJ}\; . \nn
\eea
The solution is given by
\be
 g^{(B)}_{IJ} = \frac{1}{2\pi}(\k /4\pi )^{2/3}e^{2c}\left[
                \left(\frac{1}{2\pi\r}(x^{11})^2 -x^{11}+\frac{\pi\r}{3}
                \right) S^{(1)}_{IJ}+
                \left(\frac{1}{2\pi\r}(x^{11})^2-\frac{\pi\r}{6}\right)
                S^{(2)}_{IJ}\right]\; . \label{gB}
\ee
where we have set $\orbav{g^{(B)}_{IJ}}=0$, as required. For the
calculation of the effective action we are going to need explicit
expressions for the sources $S^{(i)}$. Inserting the field strengths~\refs{Fs}
into the definition~\refs{Sdef} we find
\bea
 S^{(1)}_{\m\n} &=& e^{-c}\left[ 3e^{-2a}(D_\m CD_\n\bar{C}+
                    D_\m\bar{C}D_\n C)+\mbox{tr}F_{\m\r}^{(1)}
                    F_{\n}^{(1)\r}\right.\nn \\
                &&\qquad -\frac{1}{12}g_{\bar{\m}\bar{\n}}\left( 6e^{-4a}
                  (8|d_{rpq}C^pC^q|^2+(\bar{C}T^iC)^2)\right.\nn \\
                &&\qquad\left.\left.+12e^{-2a}|D_\m C|^2+\mbox{tr}
                    F_{\r\s}^{(1)}F^{(1)\r\s}\right)\right] \nn\\
 S^{(1)}_{AB} &=& e^{-c}\left[\frac{1}{2}e^{-2a}(8|d_{rpq}C^pC^q|^2+
                  (\bar{C}T^q_pC)^2)-\frac{1}{12}e^{2a}\mbox{tr}
                  F^{(1)}_{\r\s}F^{(1)\r\s}\right]\O_{AB} \label{Sres}\\
 S^{(2)}_{\m\n} &=& e^{-c}\left[ \mbox{tr} F_{\m\r}^{(2)}F_\n^{(2)\r}
                    -\frac{1}{12}\bar{g}_{\m\n}\mbox{tr}F_{\r\s}^{(2)}
                    F^{(2)\r\s}\right] \nn\\
 S^{(2)}_{AB} &=& e^{-c}\left[-\frac{1}{12}e^{2a}\mbox{tr}F_{\r\s}^{(2)}
                  F^{(2)\r\s}\right]\O_{AB}\; .\nn
\eea
Here $T^i$, $i=1,...,78$ are the $E_6$ generators in the fundamental
representation ${\bf 27}$. One notes that, as claimed above, both
$S^{(1)}_{AB}$ and $S^{(2)}_{AB}$ are proportional to $\O_{AB}$. 
These results, together with eq.~\refs{gB}, determine the background part
$ g^{(B)}_{IJ}$ of the metric completely.

Finally, we should discuss the structure of the three--form field
$C_{IJK}$. To do this as systematically as possible, we split
$C_{IJK}$ into three pieces as
\be
 C_{IJK} = C^{(0+1)}_{IJK}+C^{(B)}_{IJK}+\tilde{C}_{IJK}
\ee
and correspondingly
\be
 G_{IJKL} = G^{(0+1)}_{IJKL}+G^{(B)}_{IJKL}+\tilde{G}_{IJKL}
\ee
which we discuss separately. Each of these pieces will satisfy the
equation of motion and the Bianchi identity separately. The first piece,
$G^{(0+1)}_{IJKL}$, is the part which contains the actual zero mode fields
of the low energy theory, and it includes potential corrections to those zero
modes of the order $\k^{2/3}$. The second piece, $G^{(B)}_{IJKL}$,
corresponds to the components of the form which are switched on by
the gauge and gauge matter fields via the source terms in the Bianchi
identity. The last piece, $\tilde{G}_{IJKL}$, is simply the form--field part
of Witten's background which we have discussed at length in section 2 and 3.
Its nonvanishing components are $\tilde{G}_{ABCD}$ and $\tilde{G}_{ABC11}$.
Note that, while the zero mode piece $C^{(0+1)}_{IJK}$ contains zeroth order
and order $\k^{2/3}$ contributions, the background pieces $C^{(B)}_{IJK}$ and
$\tilde{C}_{IJK}$ are both of order $\k^{2/3}$.

Let us start to compute the zero mode piece $C^{(0+1)}_{IJK}$. Its two
nonvanishing components can be written as
\be
 C^{(0+1)}_{\m\n 11}=\frac{1}{6}B_{\m\n}\; ,\quad 
 C^{(0+1)}_{AB11} = \frac{1}{6}\c\left(\o_{AB} + \o'_{AB}\right)\; .
 \label{Czero1}
\ee
For the field strength we get
\be
 G^{(0+1)}_{\m\n\r 11} = 3\partial_{[\m} B_{\n\r ]}\; ,\quad
 G^{(0+1)}_{\m AB 11} = \partial_\m\c \left(\o_{AB} + \o'_{AB}\right)\; .
 \label{Gzero1}
\ee
The only modification, as compared to the zeroth order
expressions~\refs{C0mod},
is the correction $\o'_{AB}$ which has to be added in order to make
$C^{(0+1)}_{AB11}$ a zero mode of the distorted background metric. This is in
analogy to what we have discussed for the gauge matter zero mode.
A crucial difference is, however, that we now have to
deal with the full 11--dimensional metric since $C_{AB11}$ is a bulk field.
Consequently, an expansion of $\o'_{AB}$ in terms of harmonics on the
Calabi--Yau space will lead to $x^{11}$--dependent expansion coefficients.
This means that the massless part of this expansion cannot be absorbed by
a redefinition of moduli fields (which are $x^{11}$--independent), unlike
in the case of gauge matter. From the equation of motion for $C$, 
the metric~\refs{metric3}, the relations~\refs{susyeq1}--\refs{susyeq5} and
the equations~\refs{keq} we find
\be
 \o '_{AB}=\frac{\sqrt{2}}{3}\left(\cB_{AB}-\frac{1}{4}\o_{AB}\cB\right)
           +\mbox{massive terms}\; .\label{omp}
\ee
The key in arriving at this result is the observation that the
quantity $\b_A$ defined in eq.~\refs{bta} has no contributions
from massless Calabi--Yau modes. The massive terms in eq.~\refs{omp}
can be computed as well but, as argued before, will not be needed in
the following --- they are orthogonal to the zeroth-order expression
for $C_{AB11}$ and so give a vanishing contribution after integration
over the Calabi-Yau space. 

We now turn to the background $C^{(B)}_{IJK}$, which is in analogy to the
metric background $g^{(B)}_{IJ}$. It originates from the source terms in
the Bianchi identity, as has been already explained in section 2.2 in a
somewhat different context. For convenience, let us repeat some of the
essential steps here. The basic problem is to solve the equation of
motion $D_IG^{(B)IJKL}=0$ and the Bianchi identity $dG^{(B)}=0$ subject
to the boundary conditions
\bea
 \left.G^{(B)}_{\bar{I}\bar{J}\bar{K}\bar{L}}\right|_{x^{11}=0} 
   &=& -\frac{1}{4\sqrt{2}\pi}\left(\k/4\pi\right)^{2/3} 
        J^{(1)}_{\bar{I}\bar{J}\bar{K}\bar{L}} \nn \\
 \left.G^{(B)}_{\bar{I}\bar{J}\bar{K}\bar{L}}\right|_{x^{11}=\pi\r} 
   &=& \frac{1}{4\sqrt{2}\pi}\left(\k/4\pi\right)^{2/3} 
        J^{(2)}_{\bar{I}\bar{J}\bar{K}\bar{L}} \label{bdryG1}
\eea
where the sources $J^{(i)}$ are defined as
\be
 J^{(i)}_{\bar{I}\bar{J}\bar{K}\bar{L}} 
        = 6\left( {\rm tr}F^{(i)}_{[\bar{I}\bar{J}}F^{(i)}_{\bar{K}\bar{L}]}
         - \frac{1}{2}{\rm tr}R_{[\bar{I}\bar{J}}R_{\bar{K}\bar{L}]}
      \right) \; . \label{Jdef1}
\ee
A solution to this boundary value problem, for source terms $J^{(i)}$
varying slowly over scales comparable to the separation $\r$ of the
hyperplanes, is given by
\be
  C^{(B)}_{\Ib\Jb\Kb} = -\frac{1}{24\sqrt{2}\pi}\left(\k/4\pi\right)^{2/3}
       \left\{ \o^{(1)}_3 
          - (x^{11}/\pi\r)(\o^{(2)}_3+\o^{(1)}_3)
          \right\}_{\Ib\Jb\Kb} \; , \label{Cbkgd1}
\ee
where the Chern--Simons three--forms $\o_3^{(i)}$ are defined by
$J^{(i)}=d\o_3^{(i)}$ and can be expressed in terms of the Yang--Mills
and Lorentz Chern--Simons forms as in eq.~\refs{CS_def}.
Note that this approximation is well justified in the case under consideration,
since the sources are generated by low energy fields. The field strengths
$G^{(B)}=6dC^{(B)}$ are then given by
\bea
  G^{(B)}_{\Ib\Jb\Kb\Lb} &=& -\frac{1}{4\sqrt{2}\pi}\left(\k/4\pi\right)^{2/3}
      \left\{ J^{(1)} - (x^{11}/\pi\r)(J^{(2)}+J^{(1)})
      \right\}_{\Ib\Jb\Kb\Lb} \nn \\
  G^{(B)}_{\Ib\Jb\Kb 11} &=&
       -\frac{1}{4\sqrt{2}\pi^2\r}\left(\k/4\pi\right)^{2/3}
      \left(\o^{(1)}_3+\o^{(2)}_3\right)_{\Ib\Jb\Kb} \; .
\eea
The solution is a simple linear interpolation between the given values
at the two boundaries. We see that the $x^{11}$--independent component
$G^{(B)}_{\Ib\Jb\Kb 11}$ plays a r\^ole similar to the weakly coupled
Chern--Simons form and will, therefore, give rise to terms in the effective
action which are familiar from the weakly coupled case. The
$x^{11}$--dependent component $G^{(B)}_{\Ib\Jb\Kb\Lb}$ clearly has no
direct analog in the weakly coupled theory. We have seen in section 2.2
that it is needed to properly reproduce certain weakly coupled Green--Schwarz
terms in the 10--dimensional limit. Correspondingly, in the computation of
the effective action it will lead to terms which, in the weakly coupled
theory, arise from those Green--Schwarz terms,
as we will see explicitly later on.

Let us now be more specific about the nonvanishing components of
$G^{(B)}_{IJKL}$. The components $F^{(1)}_{\m b}$ and
$F^{(1)}_{BC}$ of the gauge field strength lead to a nonvanishing
source $J^{(1)}_{\m ABC}$, which affects the
components $G^{(B)}_{\m ABC}$ and $G^{(B)}_{ABC11}$. Since we do not consider
hidden sector matter, the source term at $x^{11}=\pi\r$ vanishes and
we have
\bea
 G^{(B)}_{\m ABC} &=&  -\frac{1}{4\sqrt{2}\pi}(\k /4\pi )^{2/3}
                     \left( 1-\frac{x^{11}}{\pi\r}\right)J^{(1)}_{\m ABC} \\
 G^{(B)}_{ABC11} &=& -\frac{1}{4\sqrt{2}\pi^2\r}(\k /4\pi )^{2/3}
                     \o^{(1)}_{3\, ABC}\; .
\eea
Explicitly, we find for the sources
\be
 J^{(1)}_{\m abc} = 12i\e_{abc}d_{pqr}(D_\m C)^pC^qC^r\; ,\quad
 \o^{(1)}_{3\, abc} = 4i\e_{abc}d_{pqr}C^pC^qC^r\; .
\ee
Furthermore, the square of $F^{(1)}_{\m b}$ leads to sources
$J^{(1)}_{\m\n AB}$ at $x^{11}=0$. As before, the source at $x^{11}=\pi\r$
vanishes so that
\bea
 G^{(B)}_{\m\n AB} &=&  -\frac{1}{4\sqrt{2}\pi}(\k /4\pi )^{2/3}
                     \left( 1-\frac{x^{11}}{\pi\r}\right)J^{(1)}_{\m\n AB}
 \label{bGmnAB}  \\
 G^{(B)}_{\m AB11} &=& -\frac{1}{4\sqrt{2}\pi^2\r}(\k /4\pi )^{2/3}
                     \o^{(1)}_{3\,\m AB}\label{bGmAB11}
\eea
with
\be
 J^{(1)}_{\m\n AB} = 2i\o_{AB}\left( D_\m C D_\n\bar{C} -D_\m\bar{C}D_\n C
                     \right)\; ,\quad
 \o^{(1)}_{3\,\m AB} = i\o_{AB}\left( CD_\m\bar{C} -\bar{C}D_\m C\right)\; .
\ee
Finally, the squares of the external gauge field strengths $F^{(i)}_{\m\n}$
give rise to sources $J^{(i)}_{\m\n\r\s}$ on both hyperplanes. The solution
then reads
\bea
 G^{(B)}_{\m\n\r\s} &=& -\frac{1}{4\sqrt{2}\pi}(\k /4\pi )^{2/3}
                        \left\{ J^{1}-(x^{11}/\pi\r )(J^{(2)}+J^{(1)})
                        \right\}_{\m\n\r\s}
  \label{bGmnrs} \\
 G^{(B)}_{\m\n\r 11} &=& -\frac{1}{4\sqrt{2}\pi^2\r}(\k /4\pi )^{2/3}
                         \left\{ \o_3^{(1)}+\o_3^{(2)}\right\}_{\m\n\r}
 \label{bGmnr11}
\eea
with
\be
 J^{(i)}_{\m\n\r\s} =
            6\left( \mbox{tr}F^{(i)}_{[\m\n}F^{(i)}_{\n\r ]}-\frac{1}{2}
           \mbox{tr}R_{[\m\n}R_{\n\r ]}\right)\; ,
\ee
and $J^{(i)}_{\m\n\r\s} = (d\o^{(i)}_3)_{\m\n\r\s}$. This completes our
survey of zero modes and background fields, and we are now ready to
discuss the effective action including correction terms.

\subsection{Order $\k^{2/3}$}

Let us turn to calculating the effective action at order
$\k^{2/3}$. As we have seen, background fields $G$ and gauge fields are
excited at this order and the background metric gets a
correction. Furthermore we can now consider zero modes of the gauge
fields.

By inserting the field configuration discussed in the previous subsection
into the action~\refs{LYM}, we find the contributions
\be
 S^{(1)} = 2\pi\r V\int_{M^4} \av{k_{IJ}}\left.\frac{\d S_{\rm SG}}
                      {\d g_{IJ}}\right|           
           -\frac{\pi\r V}{\k^2}\int_{M^4} \sqrt{-g}\av{\o^{AB}\o'_{AB}}
           e^{-2\hat{c}}\partial_\m\c\partial^\m\c
 \label{S1}
\ee
and
\bea
 {S^{(1)}}' &=& \frac{\pi\r V}{\k^2}\int_{M^4}\sqrt{-g}\left[ -3e^{-\hat{c}}
            D_\m CD^\m\bar{C}-\frac{3i}{\sqrt{2}}e^{-2\hat{c}}
           \left(\bar{C}D_\m C -CD_\m\bar{C}\right)
           \partial^\m\c\right.\nn \\
            &&\qquad\qquad\qquad\qquad\left. -\frac{3k^2}{4}e^{-2\hat{c}-6a}
              |d_{pqr}C^pC^q|^2-\frac{3k^2}{32}e^{-2\hat{c}-6a}
              (\bar{C}T^iC)^2\right] \nn \\
         &&-\frac{V}{8\pi\k^2}\left(\frac{\k}{4\pi}\right)^{2/3}
           \int d^4xe^{6a}\sqrt{-g}\left[ {\rm tr}(F^{(1)}_{\m\n})^2+{\rm tr}
           (F^{(2)}_{\m\n})^2\right]\; .
 \label{S1p}
\eea
to the four--dimensional effective action. One notes the appearance of
the GUT coupling constant first given in equation~\refs{couplings} above.
Here the constant $k$ is given by $k=4\sqrt{2\r}\pi (4\pi /\k )^{1/3}$
and $T^i$, $i=1,...,78$ are the $E_6$ generators in the fundamental
representation ${\bf 27}$. In order to normalize
the kinetic term for the gauge matter field $C$, we have applied the
redefinition
\be
 C^p\rightarrow \pi\sqrt{2\r}\left(\frac{4\pi}{\k}\right)^{1/3}C^p\; .
 \label{C_red}
\ee
The $R^2$ terms in the 10--dimensional boundary theories contribute higher
derivative terms at the order $\k^{2/3}$, which we have omitted in the
above expressions. An explanation about the origin and the meaning of the
various terms is in order. The first part, $S^{(1)}$, of the corrections
results from the change
in the bulk supergravity action induced by the metric distortion in
eq.~\refs{metric3} and the distortion of the components $G_{\m AB11}$ in
eq.~\refs{Gzero1}. The expression $\left.\d S_{\rm SG}/\d g_{IJ}\right|$
denotes the metric variation of the supergravity action with the zeroth
order fields inserted. It is multiplied by $\orbav{k_{IJ}}$ which, as
shown in eq.~\refs{vanish}, vanishes. The second term in
eq.~\refs{S1} originates from the part $G_{\m AB11}G^{\m AB11}$ of the
four--form kinetic term with the zero mode~\refs{Gzero1} inserted.
However, from the eqs.~\refs{omp} and \refs{vanish}, we see,
as before, that the orbifold average of the correction $\o'_{AB}$ is zero
and, therefore, this term vanishes too. Consequently, $S^{(1)}=0$
and we conclude that {\it there are
no order $\k^{2/3}$ bulk corrections to the effective action induced by the
metric distortion or the distortion of the form field moduli.} As we have
shown previously, this fact is directly related to the dropping out of massive
Calabi--Yau modes for terms linear in the distortion, and to our specific
definition of the moduli fields.

The second piece, ${S^{(1)}}'$, of the correction contains the terms
involving gauge fields. All terms in ${S^{(1)}}'$ are standard ones and
are familiar from the weakly coupled case. The kinetic term for $C$ originates
from the Yang--Mills term $\mbox{tr}{F^{(1)}}^2$ on the boundary at
$x^{11}=0$. The second term in eq.~\refs{S1p} comes from a mixing between
the zero mode $G^{(0+1)}_{\m AB11}$, eq.~\refs{Gzero1}, and the background
$G^{(B)}_{\m AB11}$, eq.~\refs{bGmAB11}. The second and the third line
of eq.~\refs{S1p} represent the scalar field potential and the Yang--Mills
action for the four--dimensional gauge fields, respectively.
 
The full bosonic effective action to the order $\k^{2/3}$ is given by the
sum of eq.~\refs{S1p} and the zeroth order action~\refs{action4}. We have
to be careful, however, about the definition of the three-form field
strength $H$ in eq.~\refs{action4}. Since we have not included the
background field $G^{(B)}_{\m\n\r 11}$ in ${S^{(1)}}'$, we should identify
$H_{\m\n\r}$ with the full component $G_{\m\n\r 11}$ of the four--form;
that is, we should define
\be
 H_{\m\n\r}\equiv G_{\m\n\r 11} = G^{(B)}_{\m\n\r 11}+G^{(0+1)}_{\m\n\r 11}\; .
\ee
The additional background $G^{(B)}_{\m\n\r 11}$ in this definition then
leads, via eq.~\refs{bGmnr11}, to the familiar nontrivial Bianchi identity
\be
  4\partial_{[\m}H_{\n\r\s]} = 
      -\frac{3}{2\sqrt{2}\pi^2\r}\left(\frac{\k}{4\pi}\right)^{2/3}
      \left[ {\rm tr}F^{(1)}_{[\m\n}F^{(1)}_{\r\s]}
       + {\rm tr}F^{(2)}_{[\m\n}F^{(2)}_{\r\s]}
       - {\rm tr}R_{[\m\n}R_{\r\s]} \right]  \label{HBI}
\ee
for $H$. If we dualize $H$ to a scalar field $\s$, we pick up additional order
$\k^{2/3}$ terms due to the source terms in this
Bianchi identity. More explicitly, we add the term
$\left(\pi\r V/6\k^2\right)\int d^4x\sqrt{-g}\s\e^{\m\n\r\s}$
$\left(\partial_\m H_{\n\r\s}-\bar{J}_{\m\n\r\s}\right)$, 
where $4\bar{J}_{\m\n\r\s}$ represents the terms on the right-hand side of
equation~\refs{HBI} to the action~\refs{action4}. Integrating out
$H_{\m\n\r}$ leads to a new term in the low-energy action, namely
\be
  {S^{(1)}}'' = \frac{V}{8\pi\k^2}\left(\frac{\k}{4\pi}\right)^{2/3}
      \int_{M^4}\sqrt{-g}\left(\sqrt{2}\s\right)\left[
                      {\rm tr}F^{(1)}\Fd^{(1)}
                      + {\rm tr}F^{(2)}\Fd^{(2)}
                      - {\rm tr}R\Rd \right]\; , \label{SYMp}
\ee
where the dual field strengths are defined as usual by
$\Fd_{\m\n}=\frac{1}{2}\e_{\m\n\r\s}F^{\r\s}$ and similarly for
$\Rd_{\m\n}$. As we will see in a moment, the gauge-field terms are
precisely those required to pair with the $\exp(6a)$ terms in eq.~\refs{S1p}
to give a chiral field in the holomorphic gauge kinetic functions. Similarly,
the ${\rm tr}R\Rd$ term will be paired by supersymmetry with the
${\rm tr}R^2$ resulting from the boundary Gauss--Bonnet terms.
Both terms will be consistently neglected since they are higher derivative.

To summarize, the effective action to order $\k^{2/3}$, after dualizing the
three--form field $H$, is given by $S=S^{(0)}+{S^{(1)}}'+{S^{(1)}}''$ with
the three parts specified in eqs.~\refs{act4_dual}, \refs{S1p} and
\refs{SYMp}. Comparison with the supergravity action~\refs{SUGRA}
shows that this implies the standard expressions
\bea
 K &=& -\ln (S+\bar{S})-3\ln (T+\bar{T}-|C|^2)\nn \\
 W &=& k\, d_{pqr}C^pC^qC^r \label{k23}\\
 f^{(1)}&=&f^{(2)}=S \nn
\eea
for the K\"ahler potential, the superpotential and the gauge kinetic
functions if we identify the fields as
\be
 S=e^{6a}+i\sqrt{2}\s\; ,\quad T = e^{\hat{c}}+i\sqrt{2}\c+\frac{1}{2}|C|^2\; .
\ee
Why, after all, did none of these functions receive modifications from
the distortion of the background as compared to the weakly coupled case?
We have argued before that all massive Calabi--Yau modes drop out
in linear order in the distortion, and contributions from massless modes
vanish for our definition of the moduli after performing the orbifold
average. The only two potential sources of corrections, therefore,
arise from massless Calabi--Yau modes on the boundary (since no
orbifold average is performed) and from the $x^{11}$--dependent background
fields $g^{(B)}_{IJ}$ and $G^{(B)}_{IJKL}$ inserted in
the bulk action. The first source of corrections cannot be seen at
the order $\k^{2/3}$, since the boundary theories are already suppressed by
this amount. Those correction will, however, become relevant at the order
$\k^{4/3}$. As for the background field $G^{(B)}$, we would need
a cross term between $G^{(B)}_{\bar{I}\bar{J}\bar{K}\bar{L}}$ and the moduli
fields in the form field kinetic term. The moduli fields, however, are
entirely contained in $G_{\bar{I}\bar{J}\bar{K}11}$ and can, therefore,
only have cross terms with the $x^{11}$--independent part
$G^{(B)}_{\bar{I}\bar{J}\bar{K}11}$ of the background. As argued before,
this part of the background is the analog of the weakly coupled
Chern--Simons term and, consequently, does not lead to any ``unconventional''
terms in the effective action. At the order $\k^{4/3}$, we will find that 
the background $G^{(B)}_{\bar{I}\bar{J}\bar{K}\bar{L}}$ comes into play.
Finally, the background $g^{(B)}_{IJ}$ for the metric appears linearly at
the order $\k^{2/3}$ and, therefore, vanishes after averaging over the
orbifold.

\subsection{Some order $\k^{4/3}$ terms}

We would now like to discuss some terms of order $\k^{4/3}$. Clearly,
we cannot compute the full effective action to that order since the
original eleven-dimensional theory is generally constructed up to terms of the
order $\k^{2/3}$ only. Additional terms in the eleven-dimensional theory,
like, for example, $R^4$ terms~\cite{R4}, can be expected to appear
at the order $\k^{4/3}$. Correspondingly, the supersymmetric background
is known to order $\k^{2/3}$ only. Consequently, for any term of order
$\k^{4/3}$ in the effective action which we want to reliably compute, we
should be able to control the effect of possible $\k^{4/3}$ terms in the
original action or of order $\k^{4/3}$ distortions of the background.
Since the boundary Yang--Mills theories are already suppressed by
$\k^{2/3}$, this can be achieved for terms in the low energy action
involving gauge fields or gauge matter fields, but it seems very hard
for terms which contain bulk fields only. In this section, we will, therefore,
compute the order $\k^{4/3}$ terms involving gauge or gauge matter fields
only.

Let us discuss the question of unknown contributions to those terms
in more detail. Clearly, unknown order $\k^{4/3}$ corrections to the
original theory involving gauge fields can occur on the boundary only.
Many of those putative terms, such as, for example, $F^4$, are, on
dimensional grounds, suppressed by powers in $\k$ larger than $4/3$.
All other terms, such as, for example, $G^2F^2$, lead to ``unconventional''
powers in $\k$, which are not integer powers of $\k^{2/3}$, and we will,
therefore, assume that they do not occur in the eleven-dimensional
theory. For the same reason these terms are probably forbidden by
supersymmetry. (Even if they did appear they would generate terms in
the low energy effective action with the same unconventional power in
$\k$ which, though lower in order than the $\k^{4/3}$ terms, would
consequently not interfere with those we calculate.)
Next, we should think about unknown order $\k^{4/3}$ corrections to the
background. Clearly, for the boundary theories those terms are irrelevant.
For the bulk theory, on the other hand, they have to involve gauge fields
in order to contribute to gauge field terms in the low energy action.
We have seen that such corrections involving gauge fields already occur at
order $\k^{2/3}$; namely, the background fields $G^{(B)}$ and $g^{(B)}$
generated by the boundary sources. Clearly, those backgrounds will have
gauge field dependent order $\k^{4/3}$ contributions as well which we
have not determined. They enter the low energy effective action linearly
at order $\k^{4/3}$. From our previous experience,
the metric pieces vanish after taking the orbifold average. This is
not quite true for the form background $G^{(B)}$. It has
however, to be paired in the action with the form field zero modes 
to contribute at order $\k^{4/3}$. This is possible only for the components
$G^{(B)}_{\m\n\r 11}$ and $G^{(B)}_{\m AB11}$ which can contract with
the zero mode fields~\refs{Gzero1} in the $G^2$ term.
In conclusion, we have identified the two $\k^{4/3}$ components
$G^{(B2)}_{\m\n\r 11}$ and $G^{(B2)}_{\m AB11}$ of the background
as the only unknown sources of $\k^{4/3}$ terms involving gauge fields.
Since they have to contract with the zero mode form field, their
contributions to the low energy action will always be proportional
to the imaginary parts $\r$, $\c$ of the moduli $S$, $T$. Other terms,
which do not contain $\r$, $\c$ cannot be affected. We will find that
our results are consistent with the contributions from those backgrounds
vanishing. This is also supported by studying the analogous problem in
the reduction to 10 dimensions~\cite{low}. For practical
purposes, we impose two further constraints on the types of terms
in the low energy action we will be calculating. These are that we will not
consider higher derivative terms and terms of mass dimension larger than six
(where we count $C$, $a$, $c$ as dimension one).

With the above remarks in mind, we use the field configuration of
section 4.2 to find the following order $\k^{4/3}$ terms with gauge
or gauge matter fields
\bea
 S^{(2)} &=& \frac{\pi\r V}{\k^2}\int_{M^4}\sqrt{-g}\left[
             \frac{3}{8}e^{-2\hat{c}}(C^2D_\m\bar{C}D^\m\bar{C}+\bar{C}^2
             D_\m CD^\m C -2|C|^2D_\m CD^\m\bar{C}) \right.\nn\\
         &&\left.\qquad\qquad\qquad\qquad -\x\a_0e^{-6a}D_\m CD^\m\bar{C}
           -\frac{i}{\sqrt{2}}\x\a_0e^{-12a}(\bar{C}D_\m C-CD_\m \bar{C})
           \partial^\m\s\right.\nn \\
         &&\qquad\qquad\qquad\qquad -\frac{k^2}{4}e^{-2\hat{c}-6a}\left(
           \frac{3}{2}e^{-\hat{c}}|d_{pqr}C^pC^qC^r|^2-e^{\hat{c}-6a}\x\a_0
           |d_{pqr}C^pC^r|^2\right)\nn \\
         &&\left.\qquad\qquad\qquad\qquad +\frac{k^2}{32}\x\a_0
            e^{-12a-\hat{c}}(\bar{C}T^iC)^2\right]\nn \\
         && -\frac{V}{8\pi\k^2}\left(\frac{\k}{4\pi}\right)^{2/3}
            \int_{M^4}\sqrt{-g}\,\x\a_0\left[ e^{\hat{c}}(\mbox{tr}
            {F^{(1)}}^2-\mbox{tr}{F^{(2)}}^2)\right.\nn \\
         && \qquad\qquad\qquad\qquad\qquad\qquad\left.
             +\sqrt{2}\c (\mbox{tr}F^{(1)}
            \tilde{F}^{(1)}-\mbox{tr}F^{(2)}\tilde{F}^{(2)})\right]\; .
         \label{S2}
\eea
As before the constant $k$ is given by $k=4\sqrt{2\r}\pi (4\pi /\k )^{1/3}$
and the gauge matter field $C$ has been redefined according to
eq.~\refs{C_red}. In addition, we have introduced $\x = \sqrt{2}\pi\r /16$.
In eq.~\refs{S2}, we have two types of terms; namely, ``conventional''
ones and those proportional to the Calabi--Yau distortion $\a_0$ defined
in eq.~\refs{a_sol}. The conventional terms arise from certain components in
$G^{(B)}_{\bar{I}\bar{J}\bar{K}11}G^{(B)\bar{I}\bar{J}\bar{K}11}$ and,
therefore, appear in the same way as the corresponding terms in the weakly
coupled case. More specifically, the terms in the first line arise from
$G^{(B)}_{\m AB11}G^{(B)\m AB11}$ and serve to complete the
$|C|^2$ piece in the  $-\ln (T+\bar{T}-|C|^2 )$ part of the K\"ahler
potential. The $|C^3|^2$ term in the third line arises from
$G^{(B)}_{ABC11}G^{(B)ABC11}$ and accounts for the $|W|^2$ part of the
scalar potential in eq.~\refs{SUGRA}.

As for the terms proportional to the Calabi--Yau distortion $\a_0$,
let us start to explain terms involving the four--dimensional gauge fields.
The terms in the second to last line of eq.~\refs{S2} represent a threshold
correction to the gauge coupling proportional to $\pm\x\a_0\mbox{Re}(T)$. 
There are two distinct sources which potentially contribute to this
threshold. The first source is just the distortion $g^{(1)}$ of the
background metric, corresponding to Witten's original
calculation~\cite{w}. Its contribution has been computed using
eqs.~\refs{CY_av1}. The reason why this contribution does not vanish in
the same way the bulk correction terms did is, simply, the aforementioned fact
that we do not average the metric distortion over the orbifold, but rather
use the boundary values of this distortion only. Note also that the
Calabi--Yau averages~\refs{CY_av1} that determined the magnitude of this
contribution to the threshold, have been determined precisely for our
definition of the moduli; that is, the definition which leads to the
standard K\"ahler
potential. This fact is essential for a reliable calculation of the
threshold. If we had not determined the K\"ahler potential to order
$\k^{2/3}$, the moduli $S$, $T$ would be unnormalized to that order,
allowing for arbitrary $\k^{2/3}$ field redefinitions. This would then
lead to an ambiguity in the threshold. In addition, there is a second
source for the threshold; namely, the background metric $g^{(B)}$,
eq.~\refs{gB}, which, from eq.~\refs{Sres}, contains the gauge field
kinetic terms. Inserted into the boundary action with the explicit
$F^2$ and $R^2$ terms in this action replaced by their internal value
using eq.~\refs{FR} they lead to a second contribution to the threshold
which has exactly the same size as the first one. There is, however, a
third contribution which arises from the expansion of the bulk curvature
term up to second order
in the metric distortion with one distortion being replaced by $g^{(1)}$
(which leads to $\a_0$) and the other one being replaced by $g^{(B)}$
(which leads to $F^2$). It turns out that this contribution exactly cancels
the second one. Finally, therefore, the threshold is entirely due to
the deformation $g^{(1)}$ of the metric which arises from the internal
gauge fields and coincides with the threshold that results from the pure
background calculation done in ref.~\cite{w}.

For a holomorphic gauge kinetic function, the gauge coupling corrections
have to be paired with terms proportional to
$\mbox{Im}(T)\mbox{tr}F\tilde{F}$, which are just the terms in the last line
of eq.~\refs{S2}. Their origin is very different from the one of the
gauge coupling corrections. They result from a component of the
$CGG$ term in the 11--dimensional action, namely from
$\e^{ABCDEF\m\n\r\s}C^{(0+1)}_{11AB}\tilde{G}_{CDEF}G^{(B)}_{\m\n\r\s}$.
Recall that $\tilde{G}_{CDEF}$ is just a component of Witten's
background which, upon contraction with the $\e$ tensor, gives rise
to the quantity $\a_0$. The zero mode piece $C^{(0+1)}_{11AB}$, given
in eq.~\refs{Czero1}, contains the field $\c$, whereas the background
field $G^{(B)}_{\m\n\r\s}$ from eq.~\refs{bGmnrs} leads to 
$\mbox{tr}F\tilde{F}$. The size of this imaginary part
is exactly what is needed for a holomorphic gauge kinetic
function. This indicates that the order $\k^{4/3}$ background field
$G^{(B2)}_{\m AB11}$ discussed in the beginning of this subsection
which could potentially modify the imaginary part of the threshold does,
in fact, not contribute.

In analogy to the real and imaginary part of the correction to the gauge
kinetic function, we have two contributions to the kinetic terms of the
gauge matter field $C$ in the second line of eq.~\refs{S2}. The first
of those terms (the analog of the real part of the gauge kinetic function)
again has two potential sources. The first source is the metric distortion
$g^{(1)}$ at $x^{11}=0$ inserted in $\sqrt{-g}\mbox{tr}{F^{(1)}}^2$. The
second source is the metric distortion $g^{(B)}$ inserted into the 
boundary action and into the second order expansion of the bulk curvature
term. In complete analogy with the threshold calculation the two latter
contributions cancel against each other and we remain with the first one
generated by $g^{(1)}$. The second term in the second line of eq.~\refs{S2},
which is the analog of the
$\mbox{Im}(T)\mbox{tr}F\tilde{F}$ piece, comes from the $CGG$ component
$\e^{\m\n\r\s ABCDEF}C^{(0+1)}_{\m\n 11}G^{(B)}_{\r\s AB}
\tilde{G}_{CDEF}$. Again, $\tilde{G}_{CDEF}$ contracts to $\a_0$.
The zero mode $C^{(0+1)}_{\m\n 11}$ in eq.~\refs{Czero1} contains the
two--form $B_{\m\n}$, which dualizes to $\s$, and $G^{(B)}_{\r\s AB}$ is the
background given in eq.~\refs{bGmnAB}, which leads to the $C\partial C$ terms.
Comparison with eq.~\refs{S1p} (first line) shows that these terms are
similar in structure to those that gave rise to the $|C|^2$ piece
in the $-\ln (T+\bar{T}-|C|^2)$ part of the K\"ahler potential, but
with $T$ replaced by $S$. Consequently, they modify the $S$ part of
the K\"ahler potential to $-\ln (S+\bar{S}-\x\a_0|C|^2)$. The potential
terms proportional to $\a_0$ in the third and fourth line of eq.~\refs{S2},
are precisely those necessary to account for this change of the K\"ahler
potential such that the superpotential remains unmodified.

To summarize, the bosonic action to order $\k^{2/3}$
in all generic fields, and to the order $\k^{4/3}$ in gauge fields and
generic gauge matter fields, excluding higher derivative terms and
terms with mass dimension larger than six is given by
$S=S^{(0)}+{S^{(1)}}'+{S^{(1)}}'' +S^{(2)}$. The various parts of this
action are defined in eqs.~\refs{act4_dual}, \refs{S1p}, \refs{SYMp}
and \refs{S2}. With the field identifications
\be
  S=e^{6a}+i\sqrt{2}\s+\frac{1}{2}\x\a_0|C|^2\; ,\quad
  T = e^{\hat{c}}+i\sqrt{2}\c+\frac{1}{2}|C|^2\; ,
\ee
we find by comparison with the supergravity action~\refs{SUGRA}
\bea
 K &=& -\ln (S+\bar{S}-\x\a_0|C|^2)-\ln (T+\bar{T}-|C|^2) \nn \\
 W &=& k\, d_{pqr}C^pC^qC^r \nn \\
 f^{(1)} &=& S+\x\a_0T \\
 f^{(2)} &=& S-\x\a_0T \nn \; ,
\eea
with $k=4\sqrt{2\r}\pi (4\pi /\k )^{1/3}$ and $\x = \sqrt{2}\pi\r /16$.
The Calabi-Yau distortion $\a_0$ is a constant,
that can be computed from eq.~\refs{a_sol} for a given Calabi--Yau space.
Clearly, at the order we are working we cannot really decide whether
$|C|^2$ in the $S$ part of the K\"ahler potential should be
inside the logarithm, but we write it in the above form for convenience.

As compared to the $\k^{2/3}$ result~\refs{k23}, we have found two new
terms, the threshold correction $\pm\x\a_0T$ to the gauge kinetic functions
and the $|C|^2$  in the $S$ part of the K\"ahler
potential. Both corrections are proportional to the deformation of
the background measured by $\a_0$. We have seen that the origin of the
$|C|^2$ term and the threshold is very similar, so that the former can be
viewed as a ``gauge matter field threshold''. Furthermore, the 
real and imaginary parts arise in a very different way for both terms.
The real part is obtained directly from the distortion of the
background metric. The imaginary part results from certain
components of the eleven-dimensional ``Chern--Simons'' term $CGG$ once the
$x^{11}$--dependent background field $G^{(B)}$ is taken into account.

\section{Conclusion}

In this paper, we have systematically derived the four-dimensional effective
action of strongly coupled heterotic string theory starting from the
eleven-dimensional effective theory as constructed by Ho\v rava and
Witten. In this derivation, we have taken Witten's background solution for
$N=1$ supersymmetry in four dimensions, good to order $\k^{2/3}$. The
solution has a distorted Calabi-Yau space as well as a non-zero
four-form. We have proven several
useful properties of this solution which were needed in the derivation
of the effective action. Specifically, we have shown that Witten's solution
can be completely expressed, in a simple way, in terms of a harmonic
two--form $\cB_{AB}$ on the internal seven--dimensional space. An explicit
solution for this two--form has been given in terms of an expansion in
harmonics of the Calabi--Yau space. Furthermore, we have demonstrated
that Witten's solution, which has been originally derived by requiring
$N=1$ supersymmetry, is indeed a solution of the equations of motion;
that is, it satisfies the linearized Einstein equation in the bulk.
A calculation of the source terms, which are needed to support the
solution on the boundaries, provided evidence for Gauss--Bonnet
$R^2$--terms in the ten-dimensional boundary actions at order
$\k^{2/3}$. Further evidence was provided by the reduction of the
theory to ten dimensions. Such terms were necessary to ensure
supersymmetry of the ten-dimensional theory. 

A feature of the theory, which we have emphasized in this paper, is that
the source terms in the Bianchi identity and the Einstein equation
provided by low energy gauge or gauge matter fields switch on
$x^{11}$--dependent components of the metric and the four-form. This
is a significant deviation from the conventional dimensional reduction
where one would take only fields independent of $x^{11}$. Further, some of
the new components would, in an ordinary dimensional reduction, be projected
out by $Z_2$--invariance. These backgrounds have been approximately
determined using a momentum expansion scheme, valid if the source terms
fluctuate on scales much larger than the radius of the orbifold. Clearly, these
backgrounds are important for any reduction of the theory to lower dimensions,
and for the derivation of the four--dimensional effective action in
particular. As an application, we have explained how the eleven-dimensional
theory can be properly reduced to its ten-dimensional limit. The
$x^{11}$--dependent backgrounds played an important r\^ole in
reproducing the Green--Schwarz terms of the ten-dimensional theory.

The central issue of this paper was the derivation of the four--dimensional
$N=1$--supersymmetric effective action. This has been done to order
$\k^{2/3}$ in all generic fields, and to order $\k^{4/3}$ in the
generic matter fields. We have excluded higher-derivative
terms and, at order $\k^{4/3}$, terms of mass dimension larger than six.
One main result is that we find the K\"ahler
potential does not receive any corrections of order $\k^{2/3}$, provided we
make an appropriate definition of the moduli fields to this order. At
order $\k^{4/3}$, an unconventional matter field correction
to the dilaton part of the K\"ahler potential appears. Altogether we find
\be
 K = - \ln \left( S+\bar{S}-\frac{\sqrt{2}\pi\r}{16}\a_0|C|^2\right)
      - 3\ln \left(T+\bar{T}-|C|^2\right) \; .
\ee
Crucial in the derivation of this result was the definition of the moduli
$S$ and $T$ in terms of the underlying eleven-dimensional geometry. The real
part of $S$, for example, should be chosen as the average Calabi-Yau volume
in order to arrive at the above K\"ahler potential. It is precisely this
definition of the moduli which we used in the further computations. We
stress that, for a meaningful computation of the threshold, it is
necessary to compute the K\"ahler potential to order $\k^{2/3}$, since,
otherwise, the field $S$, $T$ would be ambiguous to that order.

With the definition of moduli fields as explained above, we have derived
the superpotential and the gauge kinetic functions. The superpotential
remains unchanged, and is given by
\be
 W = k\, d_{pqr}C^pC^qC^r\; .
\ee
For the gauge kinetic functions we find 
\be
 f^{(1)} = S+\frac{\sqrt{2}\pi\r}{16}\a_0 T \qquad
 f^{(2)} = S-\frac{\sqrt{2}\pi\r}{16}\a_0T \; ,
\ee
where the indices $(1)$, $(2)$ refer to the gauge groups $E_6$ and $E_8$.
The threshold part of this expression, as well as the $|C|^2$ correction
to the K\"ahler potential above, are proportional to the constant
$\rho\a_0$. This is of order $\e=\k^{2/3}\rho/V^{2/3}$ ($V$ is the average
volume of the Calabi-Yau space, while $\rho$ is the orbifold radius)
and, as the dimensionless expansion parameter of Witten's solution,
measures the distortion of the Calabi-Yau space. The precise value of
$\a_0$ for a given Calabi-Yau manifold can be explicitly computed
from eq.~\refs{a_sol}. It is interesting to see how the real and
imaginary parts of the threshold arise in our M--theory
calculation.  The real part is directly related to the linear increase of
the Calabi-Yau volume along the orbifold direction due to the internal
gauge background field, in the sense first explained in ref.~\cite{w}.
We have seen that the background metric induced by the external gauge fields
can also potentially contribute to the threshold. There are, however,
two terms arising from this background, one from the boundary and the other
one from the bulk curvature term, which exactly cancel each other.
The low energy threshold is, therefore, entirely determined by the
Calabi--Yau deformation due to internal gauge fields and corresponds to
the result of ref.~\cite{w}.

The imaginary part of the threshold arises from the
eleven-dimensional $C\wedge G\wedge G$ ``Chern--Simons''-term, but with
a nontrivial background $G_{\m\n\r\s}$ inserted. This background is
required by the five-dimensional Bianchi identity with the four-dimensional
gauge fields and the four-dimensional metric as the source terms,
in much the same way that the internal Bianchi identity leads to a
distortion of the Calabi-Yau space.  The origin of the two terms in the
action which account for the $|C|^2$ correction in the K\"ahler potential is
in complete analogy with this. While the ``real'' part comes from the
distortion of the metric, the ``imaginary'' part results from
the $C\wedge G\wedge G$ term.

How are we to interpret these results as compared to the
effective low energy theory of the weakly coupled heterotic string?
To zeroth order in the expansion in $\k^2$, we certainly expect the
strongly and weakly coupled limits to give the same effective action in
four-dimensions, since nothing in the solution to this order is sensitive
to the relative sizes of the Calabi-Yau space and the orbifold. The fact
that the form of gauge kinetic functions also agrees up to order $\k^{4/3}$
with the weakly coupled result~\cite{ck,din,in}, can be
interpreted as an example of the power of holomorphy, as has been
argued by Banks and Dine~\cite{bd}. Such an argument, however, does not
apply to the K\"ahler potential. We have traced the vanishing of order
$\k^{2/3}$ corrections to the K\"ahler potential to the fact that those
correction would arise from terms linear in the background distortion.
Therefore, the part of the distortion corresponding to massive
Calabi--Yau modes drops out because of orthogonality whereas the
massless part can always be absorbed into a redefinition of the moduli.
From the generality of this argument, it is clear that the non--correction
statement for the K\"ahler potential to that order extends to non--generic
moduli as well. It is also clear that bulk corrections quadratic in the
distortion, which are of order $\k^{4/3}$, cannot vanish in general.
Therefore, at this order, one can expect the K\"ahler potential for
$S$ and $T$ to receive correction terms which depend heavily on the
distortion of the background. Those terms could potentially distinguish
the effective theory of the strongly coupled heterotic string from
its weakly coupled counterpart. Unfortunately, at the present stage,
these terms are not accessible to computation since the eleven-dimensional
theory is generally constructed up to the order $\k^{2/3}$ only.

What is the meaning of the threshold correction and the $|C|^2$
K\"ahler correction in comparison to the weakly coupled case? 
First, we should point out that the $|C|^2$ piece is compatible with
the weakly coupled scaling symmetries~\cite{bfq,hp1,ffqqv}, and can
arise as a one--loop term in the weakly coupled effective theory. There,
the imaginary part of both terms can be found by
dimensional reduction of the ten-dimensional Green--Schwarz terms. This
is well known for the imaginary part of the threshold~\cite{ck,din,in} and,
though less well known, it is true for the imaginary part of the $|C|^2$
term as well. Given that, in our context, these imaginary
parts arise from the $CGG$ term which, in turn, gives
rise to some of the weakly coupled Green--Schwarz terms upon reduction
to 10 dimensions, this is not surprising. The real part of the threshold
can be found from $R^4$ and $F^4$ terms in the 10--dimensional weakly
coupled effective theory or, more directly, from a certain large radius
limit of a string one--loop calculation. This has been explicitly
demonstrated in ref.~\cite{ns}. In the strongly coupled case,
it is obtained from the distortion of the background which, therefore, encodes
some of the string one--loop information. Similarly, the real part of the
$|C|^2$ K\"ahler correction arises from $R^4$ and $F^4$ terms in the
weakly coupled theory and represents a one--loop correction. It is
important to note that while the same threshold terms are present in the
weakly-coupled theory, the size of the corrections are quite different.
In the strongly coupled case, the correction is proportional to $\e$ and
so can be appreciable, while the analogous terms in the weakly coupled
limit are generally rather small.

In summary, we have seen that, to the order in $\k$ which we can
address at present, the form of the four--dimensional effective theories for
the weakly and strongly coupled heterotic string cannot be distinguished
from each other. We believe that this can be systematically understood
by a reduction of the eleven--dimensional theory to ten dimensions with
all backgrounds taken into account~\cite{low}. This presumably reproduces
much, or all, of the one--loop structure of the weakly coupled heterotic
string. Partial evidence for this has been given by showing that
the ten--dimensional Green--Schwarz term can be properly reproduced from
eleven dimensions.

\vspace{0.4cm}

{\bf Acknowledgments} D.~W. acknowledges with pleasure the kind hospitality
of Dieter L\"ust and Humboldt Universit\"at where part of this work was
completed. A.~L.~is supported by a fellowship from
Deutsche Forschungsgemeinschaft (DFG). A.~L.~and B.~A.~O.~are
supported in part by DOE under contract
No. DE-AC02-76-ER-03071. D.~W.~is supported in part by DOE under
contract No. DE-FG02-91ER40671.

\end{document}